\documentstyle[tighten,preprint,pre,aps,psfig,floats]{revtex}
\begin{document}
\draft
\title{Semiclassical quantization\\
using Bogomolny's quantum surface of section}
\author{M.\ R.\ Haggerty\cite{email}}
\address{W.\ K.\ Kellogg Radiation Laboratory, 106-38\\
California Institute of Technology, Pasadena, California 91125}
\date{\today}
\maketitle

\begin{abstract}
The efficacy and accuracy of Bogomolny's method of the quantum
surface of section is evaluated by applying it to the quantization of
the motion of a particle in a smooth 2-D potential. This method
defines a transfer operator $T$ in terms of classical trajectories of
one Poincar\'e crossing; knowledge of $T$ provides information about
the eigenstates of the quantum system. By using a more robust
quantization criterion than the one proposed by Bogomolny, we are
able to locate more than five hundred quantum states in both the
regular and the chaotic regimes---in most cases unambiguously---and
see no reason that the spectra could not be continued indefinitely.
The errors of the predictions are comparable in the two regimes, and
roughly constant for increasing excitation, but grow as a fraction of
the (shrinking) mean level spacing. We also show computed surface of
section wavefunctions, and present other theoretical and practical
results related to the technique.
\end{abstract}

\tableofcontents
\newpage

\narrowtext
\section{Introduction}

Semiclassical methods of quantizing certain types of Hamiltonian
systems have been known since the discovery of quantum mechanics.
Specifically, if a Hamiltonian is classically integrable (if it has
as many constants of motion as degrees of freedom) then its
trajectories are constrained to invariant tori, and EBK quantization
can be applied. Quantization occurs when the action integrated along
any closed loop on one of these tori satisfies
\[ \oint_{C_i} p\cdot dq= 2\pi\hbar (n_i+\mu_i/4)\;, \]
where $\mu_i$ is an integer that counts the number of caustics along
the trajectory. For 1-D Hamiltonians (all of which are integrable),
the tori are just the periodic orbits, and the analogous WKB method
can be applied; it yields accurate results with little effort, even
for the ground state. However, integrable systems form only a subset
of measure zero of all Hamiltonians systems; the fact that most
famous and textbook examples are of the integrable sort is because
they are easier to handle, not because nonintegrable systems are
intrinsically less interesting.

Steps towards understanding how to quantize generic, {\em
nonintegrable} systems semiclassically are more recent. The approach
that currently dominates the field is the trace formula of
Gutzwiller, which sums purely classical information about periodic
orbits into an expression for the quantum mechanical density of
states; the poles of the expression indicate quantum eigenenergies
\cite{Gut67,Gut69,Gut70,Gut71}. However, the number of periodic
orbits increases exponentially with period---faster than the
contributions of individual orbits decrease; thus the sum does not
converge absolutely. A large volume of current research is devoted to
developing clever tricks to reorder the sum in such a way that it
converges to a useful answer, but the problem is not yet
satisfactorily solved.

Recently, Bogomolny proposed an entirely different scheme to obtain
eigenstate information in the semiclassical limit by using a quantum
surface of section (SOS) \cite{Bog92}. His method, which is the topic
of this paper, will be specified precisely and discussed at length in
subsequent sections; for now, we will try to present a conceptual
overview while avoiding unnecessary details.

A quantum surface of section is akin to the classical Poincar\'e
surface of section, which has proven so useful to classical
dynamicists both practically and theoretically. A classical
Poincar\'e SOS is a surface drawn through a system's phase space; the
trajectory of interest is computed and each time that it pierces the
surface in a prespecified direction, the point where the crossing
occurred is noted. The pattern of points produced by a succession of
crossings gives information about the nature of the trajectory---for
example, whether it is periodic, quasi-periodic, or chaotic. SOS's
are most useful for systems that have two degrees of freedom; such a
system has a four dimensional phase space and a three dimensional
energy shell, but only a two dimensional surface of section (the most
convenient dimension for plotting and viewing).

Bogomolny's quantum surface of section is similarly a surface drawn
through the configuration space of the corresponding classical
Hamiltonian. Again classical trajectories are integrated from one
crossing until the next same-direction crossing of the surface. But
now, instead of only marking the points where the trajectories cross
the surface, one also notes the semiclassical phase $\exp(i S/\hbar)$
which has accumulated since the previous crossing ($S= \int\vec{p}
\cdot d\vec{q}$ is the action accumulated along the trajectory). Such
information, for all classical orbits of one Poincar\'e mapping and
some energy $E$, is summed together and projected onto the coordinate
part of the SOS into a {\em transfer operator} $T$ which will be
defined below. The projection process discards the momentum
information normally associated with a classical surface of section,
and $T$ correspondingly operates on functions of one variable fewer
than the number of degrees of freedom in the system (namely, the
Poincar\'e section's position coordinates).

Since $T$ requires only information about trajectories of one
Poincar\'e mapping, it is well defined in terms of only finite
quantities. Therefore, there are no problems at all, neither
theoretical nor practical, with divergences in Bogomolny's technique.
The $T$ matrix can be computed to arbitrary accuracy, requiring (as
we shall see) only a two dimensional numerical quadrature of
finite-time orbits. This very attractive attribute is one that is not
posessed by the Gutzwiller trace formula, which is plagued by the
exponentially growing number of periodic orbits of increasing period.

Conceptually, $T$ gives the evolution of a quantum mechanical wave
function from one intersection with the SOS to the next. In this
regard $T$ is akin to a Green function in the energy representation.
$T$ operates on functions $\vert\psi\rangle$ which live on the
coordinate part of the surface of section:
\[ \vert\psi'\rangle= T\vert\psi\rangle\;. \]
$\vert\psi\rangle$ has the value of the full quantum mechanical
wavefunction where the latter intersects the surface of section. $T$
applied to $\vert\psi\rangle$ produces, roughly speaking, the image
of $\vert\psi\rangle$ after one Poincar\'e mapping. Eigenstates of
the quantum system occur for values of adjustable parameters (which
we call $\alpha$, but could be for example $E$ or $\hbar$) for which
$T_\alpha$ has an invariant state
\begin{equation}
T_\alpha \vert\psi\rangle = \vert\psi\rangle;
\label{eq:invariant}
\end{equation}
i.e., they occur whenever $T_\alpha$ has an eigenvalue that is equal
to unity. So to find the eigenstates of a quantum mechanical system,
one computes $T$, diagonalizes it to find its eigenvalues, and plots
those eigenvalues in the complex plane for a range of $\alpha$.
Whenever one of the eigenvalues crosses through $1$, then {\em at the
corresponding set of parameters $\alpha$, the quantum mechanical
system is predicted to have an eigenstate}.

We know of four other calculations to date that use Bogomolny's
technique. Lauritzen \cite{Lau92}, by resorting to a stationary phase
integral, showed that Bogomolny's quantization condition
(\ref{eq:invariant}) reduces to EBK quantization for integrable
systems in general, and the rectangular billiard in particular.
Bogomolny and Carioli \cite{BogCar93} applied the method to a
``surface of constant negative curvature'' with vanishing potential
energy; this is a billiard-type chaotic system whose orbits can also
be written explicitly.

Szeredi, Lefebvre, and Goodings \cite{SzeLefGoo93} used the quantum
SOS in their study of the wedge billiard, a scalable system bounded
on two sides by straight hard walls and confined in the open
direction by a uniform downward gravity-like force. This system has
four types of orbits of one Poincar\'e mapping, which can be written
down; they summed these orbits into a $T$ matrix in a basis of
position-space cells and were able to reproduce the first twenty
quantum eigenvalues with an average RMS error of 6.5\% of the mean
level spacing.

Finally, Boasman in his thesis \cite{Boa92} thoroughly investigates,
in a largely analytic way, the asymptotic accuracy that Bogomolny's
method achieves for billiard problems, and supports his predictions
with evidence from numerical calculations.

Each of the previous calculations were restricted to non-generic
systems---integrable systems or billiards (or integrable billiards).
There is, of course, a reason for preferring billiards: they are
scalable systems whose classical trajectories are the same regardless
of energy, many chaotic billiards are known, and, most importantly,
one can write down explicit formulas for the classical trajectories
connecting any two points on the surface of section. On the other
hand, billiards are thought to have different convergence properties
than smooth potentials \cite{Sch94}. Moreover, smooth
potentials---not billiards---are the kind typically encountered in
models of natural systems, so it is interesting to know how well they
can be handled with new methods.

Therefore, we chose to undertake our research in this more
challenging laboratory---the smooth Hamiltonian system. The
centerpiece of this paper is a computational application of
Bogomolny's method to the Nelson$_2$\ potential (see
Appendix~\ref{sec:rescaling}), a smooth, bounded, nonlinear
oscillator with Hamiltonian
\[ H= {1\over2}\left( p_x^2 + p_y^2 \right)+
{1\over2}\omega^2 x^2+
{1\over2}\left( y- {1\over2} x^2 \right)^2\;. \]
The system is non-scalable and has a rich periodic orbit structure
\cite{BarDav87}. For low energies, it approaches a 2-D anisotropic
harmonic oscillator and is predominantly regular; as the energy is
increased, the degree of chaos increases and eventually dominates the
phase space. We will present computations in both regimes to
illuminate the similarities and differences.

This paper is organized as follows: Section~\ref{sec:Bogomolny}
outlines the idea of Bogomolny that is the subject of this paper.
Section~\ref{sec:Computing} develops various expressions for $T$
which offer a somewhat different perspective on its operation, and
which translate directly into a quadrature algorithm for computing
$T$ in a non-billiard system. Section~\ref{sec:symmetries} shows how
to take advantage of a mirror symmetry when computing $T$.
Section~\ref{sec:O(n)} estimates the effort needed to apply
Bogomolny's method, as compared with traditional methods.
Section~\ref{sec:eigenclassicities} explains a trick which enables
Bogomolny's theory to be verified with less numerical effort than a
naive approach would require. Section~\ref{sec:exact} discusses the
nature of eigenclassicity problems in general, and provides details
of how the exact eigenclassicity spectrum was calculated for the
Nelson$_2$ system. Section~\ref{sec:model} introduces the model
system to which we applied Bogomolny's method.
Section~\ref{sec:SCspectrum} gives some of the behind-the-scenes
details about our implementation of the semiclassical computation.
Section~\ref{sec:eigenvalues} qualitatively describes the behavior of
the eigenvalues of the $T$ operator. Section~\ref{sec:accuracy}
presents the eigenclassicity spectra produced by Bogomolny's method,
and compares them to exact spectra in both the regular and the
chaotic regime. Finally, section~\ref{sec:SOSwavefunctions} tells how
the surface of section wavefunctions can be obtained from the theory,
and makes some comments about how well they are predicted.

A comment about nomenclature: we will be dealing with two related but
distinct eigenproblems---Bogomolny's condition on the $T_\alpha$
operator (Eq.~(\ref{eq:invariant})) and the time independent
Schr\"odinger equation for the full quantum mechanical system. In
order to reduce confusion, we assume the following naming convention:
the terms {\em eigenvalues} and {\em surface of section
eigenfunctions} in this paper always refer to quantities obtained
from diagonalizing the $T_\alpha$ operator. It should be kept in mind
that $T_\alpha$ and its eigenvalues can be computed for any choice of
parameters $\alpha$, whether or not an eigenstate of the quantum
system exists for those parameters. The words {\em eigenenergies},
{\em eigenclassicities} (explained below), and {\em eigenstates} all
refer to energy eigenstates of the full quantum mechanical
Hamiltonian. These eigenstates only exist for special values of the
parameters $\alpha$---in fact, only those for which $T$ has a unit
{\em eigenvalue}, according to Bogomolny's theory.

\section{Theory}
\label{sec:Theory}

\subsection{The results of Bogomolny}
\label{sec:Bogomolny}

\begin{figure}
\centerline{
\psfig{file=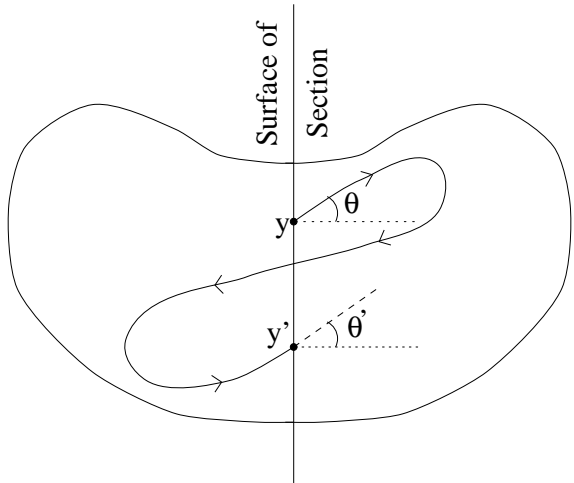,width=7.5cm}
}
\caption[Orbits which contribute to $T$]{Orbits which contribute to
the semiclassical transfer operator. In Bogomolny's construction, the
trajectories which contribute are those of one complete Poincar\'e
section: they are integrated from one intersection with the $y$-axis
until the next intersection that occurs in the same direction.}
\label{fig:trajs}
\end{figure}

Bogomolny \cite{Bog92} gives an expression for the transfer operator
$T$ for systems of any dimension and using any surface of section.
Our interests are narrower, since the systems for which we will be
doing computations are symmetric about the $y$-axis, and the $y$-axis
will be used as the surface of section. In this case, the expression
for $T$ can be written
\widetext
\begin{equation}
\langle y'\vert T\vert y\rangle=
\frac{1}{(2\pi i \hbar)^{1/2}}
\sum_{\rm cl.tr.}
\left\vert\partial^2 S(y',y)\over \partial y\partial
y'\right\vert^{1/2}
e^{-i\pi\nu/2} \exp i \frac{S(y',y)}{\hbar}\;.
\label{eq:Tsum}
\end{equation}
\narrowtext
$T$ can be calculated for any chosen value $E$ of energy. The sum is
over all classical trajectories which have that energy and start at
$(0,y)$ and end at $(0,y')$ one Poincar\'e mapping later (that is,
with no intervening same-direction piercings of the SOS) (See
Fig.~\ref{fig:trajs}).
$S(y',y) \equiv \int_y^{y'} \vec{p}\cdot d\vec{q}$ is the classical
action along the trajectory considered. The second derivative of $S$
gives the degree of focusing of nearby trajectories onto the current
trajectory. The focusing switches sign each time that there is a
perfect re-convergence of nearby trajectories, which would lead to a
branch cut ambiguity when its square root is taken; therefore, its
absolute value is taken and the phase is put in separately through
the Maslov index $\nu$, which counts the number of sign changes in
the focusing (note: its role is a bit subtler in higher dimensions;
see Ref.~\cite{Rob91}).

Bogomolny's construction of the $T$ operator begins by dividing the
allowed portion of phase space into two subregions ``1'' and ``2,''
one on either side of the surface of section. In each half, a Green
function $G_{1,2}$ is constructed which (i)~obeys Schr\"odinger's
equation in that subregion, (ii)~is arbitrary on the SOS, and
(iii)~obeys the same boundary conditions as the true wavefunctions on
the remainder of the boundary of that subregion. The next step is to
write the wavefunctions in terms of the Green functions and a source
function $\psi_{1,2}(y)$ on the surface of section:
\[ \Psi_{1,2}(x, y)=
\int_\Sigma dy'\, G_{1,2}(x,y;y';E) \psi_{1,2}(y')\;. \]
This equation, plus the demand that $\Psi_1$ and $\Psi_2$ match on
the SOS and satisfy the Schr\"odinger equation in region 1 or 2,
respectively, lead to the self-consistency requirement that
$\psi_1(y)$ satisfy
\begin{equation}
\int_\Sigma dy' \, \tilde{G}(x,y;y';E) \psi_1(y') = 0
\label{eq:Gconsist}
\end{equation}
where
\widetext
\begin{eqnarray*}
\tilde{G}(x,y;y';E) & = & \frac{\hbar^2}{2}\int_\Sigma dy' \\
& & \times \left( G_1(0,y;y';E) \frac{\partial}{\partial n}
G_2(x,y;y';E)-
G_2(x,y;y';E) \frac{\partial}{\partial n} G_1(0,y;y';E)
\right)
\end{eqnarray*}
\narrowtext
and $n$ is the outward-directed normal at point $(0,y)$.

For points on the surface of section, it is straightforward to write
the expression for $\tilde{G}$ in the semiclassical limit in terms of
a sum over classical trajectories of one Poincar\'e mapping:
\begin{eqnarray*}
\tilde{G}(y'';y';E) & = &
\sum_{\rm cl.tr.}
\frac{1}{i\hbar(2\pi i\hbar)^{1/2}}
\left\vert
\frac{1}{\vert p''\vert\, \vert p'\vert}
\frac{\partial^2 S}{\partial y'' \partial y'}
\right\vert^{1/2} \\
& & \times \exp \left(
\frac{i}{\hbar}
S(y'',y';E) - i\frac{\pi}{2} \nu \right)\;.
\end{eqnarray*}
A few more unilluminating steps transform the consistency condition
(\ref{eq:Gconsist}) into
\[ \det (1-T)= 0\;, \]
where $T$ is given by Eq.~(\ref{eq:Tsum}).

Bogomolny also gives some of the properties of the $T$ operator, and
proves them in the classical limit $\hbar \rightarrow 0$. His most
important of these subsidiary claims is that in the limit $\hbar
\rightarrow 0$, the $T$ operator is unitary. This is, to be sure, a
strange sort of unitarity, in light of his other claim that the
dimension of $T$ varies smoothly with parameters (such as energy);
specifically, he says that
\begin{equation}
\dim T_\alpha=
\frac{\text{volume of allowed region on Poincar\'e surface}}
{(2\pi\hbar)}\;.
\label{eq:Tdimension}
\end{equation}
The mechanism by which these two phenomena coexist will be examined
in detail in the context of our numerical experiment
(Sec.~\ref{sec:eigenvalues}).

There are other interesting subjects covered in Bogomolny's paper,
such as the relationship between $T$ and the Selberg zeta function,
and his prescription for computing full quantum mechanical
eigenfunctions; we will not address those challenges in this paper,
beyond presenting computations of {\em surface of section}
wavefunctions predicted by the theory and comparing them to the exact
SOS wavefunctions.

\subsection{Computing the $T$ operator: Avoiding the shooting
problem}
\label{sec:Computing}

Each of the coordinate-space matrix elements of $T$ in expression
(\ref{eq:Tsum}) above is a sum over classical trajectories of energy
$E$ which go from $(0,y)$ to $(0,y')$ in one Poincar\'e mapping. But
to find these trajectories, it would be necessary to find all values
of the initial momentum $\vec{p}= (p\cos\theta,p\sin\theta)$ that
cause a trajectory launched from $y$ to next intersect the SOS at
$y'$. Even though the momentum magnitude $p= \sqrt{2(E-V(0,y))}$ is
fixed by the choice of energy, one would still have to solve a
shooting problem---a (numerical) search in $\theta$ space to find the
launching angles which cause the particle to end up at $y'$.

But this can be avoided. Consider: any properly chosen surface of
section has the property that every trajectory eventually pierces it.
As a function of initial conditions (on the SOS), call the next
crossing point $Y'(y,\theta,E)$. It follows that {\em every}
trajectory of the appropriate energy contributes to $T$; if it starts
at $(0,y)$ with angle $\theta$, for example, it contributes to
$\langle Y'(y,\theta,E)\vert T\vert y\rangle$. This observation
suggests that we transform (\ref{eq:Tsum}) from a sum over endpoints
into an integral over initial conditions.

Executing the desired transformation is possible and indeed
straightforward. First we write a more useful expression for the
partial derivative which appears in Eq.~(\ref{eq:Tsum}), being
explicit about which variables are held constant:
\begin{eqnarray*}
\frac{\partial^2S(y',y)}{\partial y\partial y'} & \equiv &
\left[ {\partial\over\partial y'}
\left( \partial S(y',y)\over\partial y \right)_{y'E\Sigma}
\right]_{yE\Sigma} \\
& = & \left[ {\partial\over\partial y'} (-p_y) \right]_{yE\Sigma} \\
& = & -\left[ \partial y'\over\partial p_y\right]_{yE\Sigma}^{-1}\;.
\end{eqnarray*}
Here we use a subscript of ``$\Sigma$'' to remind ourselves that the
surface of section is meant to be fixed during the
differentiation---in our case, $x=x'=0$. The second line follows from
the well-known identity $\left(\partial S/\partial y\right)_{y'E}=
-p_y$.

As a function of initial conditions $y$ and $\theta$,
\begin{eqnarray*}
\left[ \partial y'\over\partial p_y \right]_{yE\Sigma} & = &
\left[ \partial Y'(y,\theta,E)\over\partial p_y \right]_{yE\Sigma}\\
& = &
\left[ \partial Y'\over\partial\theta\right]_{yE\Sigma}
\left[ \partial\theta\over\partial p_y\right]_{yE\Sigma}\\
& = &
{1\over p_x} \left[ \partial
Y'\over\partial\theta\right]_{yE\Sigma}\;.
\end{eqnarray*}
Substituting into Eq.~(\ref{eq:Tsum}) and integrating out the basis
states on the LHS, we have
\widetext
\begin{equation}
T= {1\over(2\pi i\hbar)^{1/2}} \int dy\int dy'\sum_{\rm cl.tr.}
\vert p_x \vert^{1/2}
\left\vert \partial Y'\over\partial \theta
\right\vert_{yE\Sigma}^{-1/2}
e^{iS/\hbar-i\pi\nu/2} \vert y'\rangle\langle y\vert\;.
\label{eq:T2}
\end{equation}
Now we notice that the sum over classical trajectories that go from
$y$ to $y'$ in one Poincar\'e mapping is equivalent to a sum over the
discrete values $\theta_i$ that solve the shooting problem $y'=
Y'(y,\theta_i,E)$ for the present values of $y$, $y'$, and $E$.
Schematically, we express that statement with the following
equalities, which hold no matter what expression is inserted in the
braces:
\begin{eqnarray*}
\sum_{\rm cl.tr.} \biggl\{\ldots\biggl\} & = & \int d\theta\sum_i
\delta(\theta- \theta_i)
\left. \biggl\{\ldots\biggl\} \right|_{\theta} \\
& = & \int d\theta
\left\vert \partial Y'(y,\theta,E) \over\partial \theta
\right\vert_{yE\Sigma}
\delta(y' - Y'(y, \theta, E)) \left. \biggl\{\ldots\biggl\}
\right|_{\theta}\;.
\end{eqnarray*}
\narrowtext
When we apply this identity to Eq.~(\ref{eq:T2}), the
$\delta$-function allows us to do the $y'$ integral immediately:
\begin{eqnarray}
T & = & {1\over(2\pi i\hbar)^{1/2}} \int dy \int d\theta \,
\vert p_x\vert^{1/2}
\left\vert \partial Y'(y,\theta,E) \over\partial \theta
\right\vert_{yE\Sigma}^{1/2}
\nonumber \\
& & \times \exp \left( i \frac{S(y,\theta)}{\hbar} - i \frac{\pi}{2}
\nu \right)
\vert Y'(y, \theta, E) \rangle\langle y \vert\;.
\label{eq:Tintegral}
\end{eqnarray}
The result is an expression for $T$ which can be evaluated without
solving any shooting problems---a reduction of numerical effort.
Moreover, this expression more closely represents our intuitive
picture of the effect of $T$ than Eq.~(\ref{eq:Tsum}); that is, when
$T$ is applied to an initial surface of section wavefunction
$\vert\psi\rangle$,
\begin{enumerate}
\item it breaks up $\vert\psi\rangle$ into its components at each
position $y$;
\item each of these components becomes an ensemble of classical
particles, launched in all directions $\theta$;
\item the particles follow the classical equations of motion
(accumulating quantum mechanical phase as they go) until they hit the
surface of section again;
\item the phases of the particles are summed together (with a
weighting factor) to yield the new SOS wavefunction
$\vert\psi'\rangle$.
\end{enumerate}
It is Eq.~(\ref{eq:Tintegral}) which formed the starting point for
our numerical work.

Note that the partial derivative appearing in
Eq.~(\ref{eq:Tintegral}) is not computed directly, but rather from
elements of the linearized tangent matrix, which can be computed
efficiently using techniques similar to those described by Eckhardt
and Wintgen \cite{EckWin91} for computing the monodromy matrix. (The
slight difference is that the monodromy (``once around'') matrix only
applies to periodic orbits, whereas we need to calculate stabilities
at arbitrary times on non-periodic orbits.)

\subsection{Removing symmetries}
\label{sec:symmetries}

\begin{figure}
\centerline{
\psfig{file=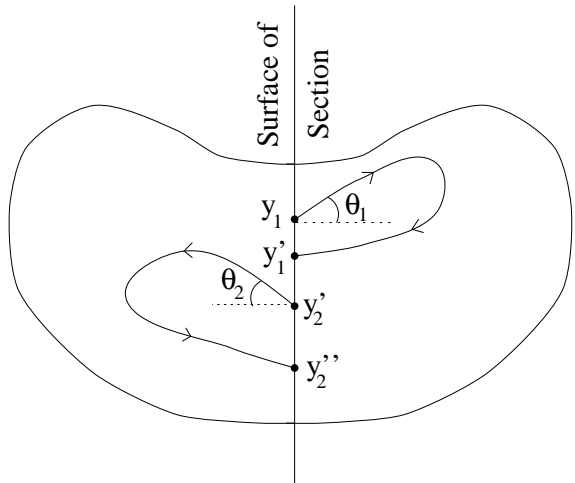,width=7.5cm}
}
\caption[Half-Poincar\'e mapping trajectories that contribute to
$T_1$ and $T_2$]{Half-Poincar\'e mapping trajectories which
contribute to $T_1$ and $T_2$. In the text it is shown that, to
within a stationary phase approximation, the transfer operator $T$
can be written as the product of $T_1$ and $T_2$, each of which is a
sum over ``half-Poincar\'e mapping'' trajectories on one side or the
other of the SOS, such as those drawn here. The stationary phase
approximation tells us that the strongest contributions to the
product will come from pairs of half trajectories that join smoothly
at the surface of section into a full-mapping trajectory.}
\label{fig:halftrajs}
\end{figure}

Remember that the $T$ operator as defined above gives the evolution
of a SOS wavefunction from one crossing of the SOS to the next {\em
same-direction} crossing. But one might think that it would be also
possible to write $T$ as the composition of two operators: a $T_1$,
which performs the evolution to the first crossing of the surface of
section (which is in the ``{\em wrong}'' direction), followed by a
$T_2$, which performs the second half of the evolution (to the second
crossing, which is the first ``{\em proper},'' same-direction
crossing; see Fig.~\ref{fig:halftrajs}).
The proof of this fact is the subject of the present section;
effectively, we need to unravel the last part of Bogomolny's
derivation of the $T$ operator.

The two operators $T_1$ and $T_2$ are defined by expressions exactly
equivalent to Eq.~(\ref{eq:Tsum}), except that their sums are over
half-Poincar\'e trajectories going from $y_1 \rightarrow y'_1$ and
$y'_2 \rightarrow y''_2$, respectively (see
Fig.~\ref{fig:halftrajs}). We wish to show that the product $T_2 T_1$
is equal to $T$. The product contains a $\delta$-function $\langle
y'_2\vert y'_1\rangle$, which allows us to perform one of the
integrals immediately, yielding
\widetext
\begin{eqnarray}
T_2 T_1 & = & {1\over 2\pi i\hbar} \int dy'' \int dy' \int dy
\sum_{y' \rightarrow y''} \sum_{y \rightarrow y'}
\left\vert\frac{\partial^2 S_2(y'',y')}{\partial y'\partial y''}
\right\vert^{1/2}
\left\vert\frac{\partial^2 S_1(y',y)}{\partial y\partial
y'}\right\vert^{1/2}
\nonumber \\
& & \times e^{-i\pi(\nu_2 + \nu_1)/2} \exp \frac{i}{\hbar}
\left( S_2(y'',y')+ S_1(y',y) \right) \, \vert y''\rangle\langle
y\vert\;.
\label{eq:T2T1}
\end{eqnarray}
\narrowtext
We next do the $y'$ integration using the stationary phase
approximation; the only significant contribution is when
\begin{eqnarray}
0 & = & {\partial\over\partial y'} \left[ S_2(y'',y')+ S_1(y',y)
\right]_{y,y'',E}
\label{eq:S1S2cond} \\
& = & \left[ -p'_{2,y}+ p'_{1,y} \right] \nonumber
\end{eqnarray}
which requires that the final momentum of the first part of the
trajectory equal the initial momentum of the second part---the
trajectories must join smoothly. At those points, the stationary
phase approximation gives an additional factor of
\[ \left| \frac{\partial^2}{\partial y'^2}
\left[ S_2(y'',y') + S_1(y',y) \right]_{y,y'',E}
\right|^{-1/2}
\]
in the integrand of Eq.~(\ref{eq:T2T1}).

It remains only to show that the new combined prefactor of
Eq.~(\ref{eq:T2T1}) matches that of Eq.~(\ref{eq:Tsum}); i.e., that
\begin{equation}
\frac{
\frac{\partial^2 S_1}{\partial y\partial y'}
\frac{\partial^2 S_2}{\partial y''\partial y'}}
{
\frac{\partial^2 S_1}{\partial y'^2}+
\frac{\partial^2 S_2}{\partial y'^2}}
\stackrel{{\textstyle ?}}{=} \pm
\frac{\partial^2 S(y'',y)}
{\partial y\partial y''}\;.
\label{eq:prefac}
\end{equation}
To this end we define the function $Y'(y,y'')$ which gives the values
of $y'$ for classical trajectories smoothly connecting $y$ to $y''$
in one Poincar\'e mapping. We then note that Eq.~(\ref{eq:S1S2cond})
is valid for any values of $y$ and $y''$ as long as we evaluate it at
$y' = Y'(y,y'')$, and so we write that equation's derivatives with
respect to $y$ and $y''$:
\begin{equation}
0= \left(
\frac{\partial^2 S_1}{\partial y'^2}+
\frac{\partial^2 S_2}{\partial y'^2} \right)
\frac{\partial Y'}{\partial y}+
\frac{\partial^2 S_1}{\partial y\partial y'}
\label{eq:Sdy1}
\end{equation}
\begin{equation}
0 = \left(
\frac{\partial^2 S_1}{\partial y'^2}+
\frac{\partial^2 S_2}{\partial y'^2} \right)
\frac{\partial Y'}{\partial y''}+
\frac{\partial^2 S_2}{\partial y''\partial y'}\;.
\label{eq:Sdy2}
\end{equation}
In the current nomenclature, the action that enters the expression
for $T$ is
\[ S(y'',y) = S_2(y'',Y'(y,y'')) + S_1(Y'(y,y''),y)\;; \]
we will need its second partial derivative, which we can obtain using
the chain rule and Eq.~(\ref{eq:S1S2cond}):
\widetext
\begin{equation}
\frac{\partial^2 S(y'',y)}{\partial y\partial y''}=
\frac{\partial^2 S_1}{\partial y\partial y'}
\frac{\partial Y'}{\partial y''}+
\frac{\partial^2 S_2}{\partial y''\partial y'}
\frac{\partial Y'}{\partial y}+
\left(
\frac{\partial^2 S_1}{\partial y'^2}+
\frac{\partial^2 S_2}{\partial y'^2} \right)
\frac{\partial Y'}{\partial y}
\frac{\partial Y'}{\partial y''}\;.
\label{eq:Sdd}
\end{equation}
\narrowtext
Using Eqs.~(\ref{eq:S1S2cond}), (\ref{eq:Sdy1}), (\ref{eq:Sdy2}), and
(\ref{eq:Sdd}), it is trivial to establish that the equality holds in
(\ref{eq:prefac}) when the minus sign is chosen. Therefore we have
established that, to within a stationary phase approximation,
\[ T \stackrel{\text{SPA}}{=} T_2 T_1\;.\]
Thus $T$ can be decomposed into ``half-Poincar\'e mapping operators''
$T_1$ and $T_2$, as we hoped.

There will be a numerical efficiency gain from using this
decomposition for any chaotic potential, regardless of symmetry, for
the following reason. The computation of $T$ requires doing an
integral over the initial conditions $y$ and $\theta$. The integrand,
however, involves functions such as $Y'(y,\theta)$, which is the
point that a trajectory next intersects the SOS. To get better than
Monte-Carlo quality convergence of the integral, these functions must
be sampled on a fine enough mesh that their variation as a function
of initial conditions is sampled. In a chaotic regime, where nearby
trajectories diverge exponentially in time, the divergence of nearby
half trajectories will be roughly the square root of the divergence
of full trajectories, and so, roughly, only the square root of the
number of mesh points will need to be used. Thus two coarser meshes
of half-length classical trajectories will be adequate to compute
$T_1$ and $T_2$, and then those operators (in the form of matrices)
can be multiplied to yield $T$. In fact, we suspect that the product
$T_2 T_1$ will yield even {\em better} estimates of the eigenvalues
of the system due to the fact that it is ``one stationary phase
approximation closer'' to the exact Feynman path integral
underpinning the semiclassical approximations.

We use the $T_2 T_1$ approach in our numerical computations below. In
our case we realize an even more significant increase in numerical
efficiency when we use the $T_2 T_1$ approach: because our potential
is symmetric with respect to reflection about the surface of section,
$T_1 \equiv T_2$. Therefore, in addition to the less dense mesh of
trajectories that need to be calculated, the second half of the
trajectories need {\em never} be calculated. Moreover there is no
need to multiply $T_1 \cdot T_1$; our criterion that $T$ have an
eigenvalue of $1$ is equivalent to the requirement that $T_1$ have an
eigenvalue of $+1$ {\em or $-1$}. Analogously, Bogomolny's condition
$\det(1-T)=0$ would become
\[ \det(1-T^2)= \det(1-T)\det(1+T)=0\;. \]
The sign of the eigenvalue tells us the parity of the associated
eigenstate of the system with respect to reflection about the SOS.

Formally this reduction to the fundamental domain is equivalent to
solving the half-domain problem with two different boundary
conditions: first with a soft wall at the SOS, and second with a hard
wall. The latter case is the one that produces odd-parity
eigenstates, as follows: each trajectory has one reflection from the
wall, and thus an additional phase of $\pi$ appears through its
Maslov index; this makes $T_{\text{hard}}= (-1)T_1$. In this picture
quantization occurs when $T_{\text{hard}}$ has an eigenvalue of $1$,
so, as above, these odd-parity states occur when $T_1$ has an
eigenvalue of $(-1)$.

Throughout the rest of the paper, we use the desymmetrized transfer
operator $T_1$ in our computations, and we drop the subscript.

It is interesting to comment that another stationary phase
approximation, similar to the one connecting $T$ and $T_2 T_1$, would
produce the Gutzwiller periodic orbit formula, the better-known
device for semiclassically quantizing chaotic systems. From this
vantage point it is easy to conjecture that the transfer matrix
approach, which is ``one stationary phase approximation closer'' to
Feynman path integration than the periodic orbit formula, will yield
correspondingly better estimates of quantum properties of the system
than the trace formula for a comparable amount of effort or a
comparable number of input classical trajectories. Unfortunately, the
implementations of the two methods differ so completely that
comparisons based on ``equal effort'' will be tricky and this
conjecture will not be tested in the present paper.

\subsection{Algorithmic complexity of method}
\label{sec:O(n)}

The algorithm that needs to be followed to compute a system's
spectrum follows from Eq.~(\ref{eq:Tintegral}). We now give a crude
estimate of the computational effort required to get the first $N$
eigenstates of a $d$-degree of freedom system which has instability
exponent $\lambda$---more precisely, we give the scaling of the
effort with those quantities.

In character with the rest of this section, we will not attempt to
give a precise definition of $\lambda$, except to say that it should
measure the ``typical'' separation of two nearby orbits during one
Poincar\'e mapping, as follows:
\[ \vert\vec{y}'_2 - \vec{y}'_1\vert \sim
e^\lambda \vert\vec{y}_2 - \vec{y}_1\vert\;. \]
We ignore the common situation that the degree of classical chaos
varies with excitation number $N$ because the nature of this
interdependence is very system-specific.

The $N$th excited state has a de~Broglie wavelength which is
$\lesssim O(N^{-1/d})$, the estimate coming from counting the number
of nodes that would fit in a container with rigid walls. Computing
$T$ requires that enough classical trajectories be calculated to
capture the dynamics of the full energy shell with a resolution
comparable to the de~Broglie wavelength of the $N$th state. Thus if
the trajectories are started from a mesh of initial conditions with
spacings in positions and momenta proportional to $\Delta$, we need
\[ O(\Delta \cdot e^\lambda) \lesssim O(N^{-1/d}) \]
so
\[ \Delta \lesssim O(N^{-1/d} e^{-\lambda})\;. \]
The mesh needs to include all trajectories of energy $E$ that start
on the SOS, a surface of dimension $(2d-2)$. This is thus also the
dimension of the mesh of initial conditions, so the number of
classical trajectories that need to be calculated is
\[ O(\Delta^{-2(d-1)}) \sim
O(N^{2(1-1/d)} e^{2\lambda(d-1)})\;. \]

Each trajectory must, in general, be integrated numerically. The
number of time steps necessary depends on the details of the
potential; a reasonable estimate is that it also scales like the
reciprocal of the mesh spacing, $O(\Delta^{-1})$. Thus the
computational effort of computing the necessary trajectories is the
number of trajectories times the number of time steps per trajectory,
or
\[ O(\Delta^{-(2d - 1)}) \sim
O(N^{2-1/d} e^{\lambda(2d - 1)})\;. \]
(It will be the case that the effort of updating the phase space
vector and stability matrix scale as a small constant power of $d$,
but that factor is negligible compared with the other contributions.)

Then the $T$ operator must be constructed from the information about
the trajectories. $T$ operates on functions on the spatial part of
the surface of section; these functions have $(d-1)$ dimensions---one
fewer than eigenstates of the full quantum system. In practice, the
matrix elements $\langle n_1\vert T_\alpha\vert n_2\rangle$ will be
calculated in some basis fine enough to capture details the size of
the de~Broglie wavelength, in $(d-1)$ dimensions---that requires
$\dim T \sim O(N^{1 - 1/d})$ basis states, so that $T$ has the square
of that or $O(N^{2(1 - 1/d)})$ matrix elements. If $T$ is to be
calculated in a generic basis, then each of its matrix elements needs
to be updated for each trajectory; a job of complexity
$O(N^{4(1-1/d)}) e^{2\lambda (d-1)}$. We can reduce this if we choose
trajectories and basis sets more carefully.

If the mesh of initial conditions is rectangular, the job becomes
somewhat easier because we can compute
\[ T \vert{\vec{y}}\rangle \]
for each initial $\vec{y}$ as an integral over initial momenta, and
only then sum the whole row into the $T$ matrix; this optimization
reduces the complexity to $O(N^{3(1-1/d)} e^{2\lambda(d-1)})$.

Even better is to choose to compute $T$ in a position basis (that is,
positions covering the surface of section). In this case, a
particular trajectory only contributes to a single matrix element of
$T$ (or at most a few, depending on the rounding scheme). Thus
updating the $T$ matrix need not take more than constant time for
each trajectory, and this part of the algorithm is reduced from being
fatally expensive to being almost incidental---only $O(N^{2(1-1/d)}
e^{2\lambda(d-1)})$ \cite{footnote}.

Next the $T$ matrix needs to be diagonalized, with effort that goes
with the cube of the size of the matrix, $O(N^{3(1-1/d)})$. In this
step Bogomolny's method has an advantage over a brute-force
diagonalization of the Hamiltonian, which requires a matrix with size
$O(N)$ and effort $O(N^3)$.

Finally, $\alpha$ must be scanned to find parameter values that yield
eigenstates. This procedure requires $O(N)$ repetitions of each of
the above steps.

A grand total of the computational effort required to apply
Bogomolny's method incorporates all of the above estimates:
\widetext
\begin{eqnarray}
\text{effort} & \sim &
O([
\underbrace{N^{2-1/d} e^{\lambda(2d - 1)}}_{\text{calc.
trajectories}}+
\underbrace{N^{2(1-1/d)} e^{2\lambda (d-1)}}_{\text{update $T$}}+
\underbrace{N^{3(1-1/d)}}_{\text{diagonalize $T$}}
] \cdot \underbrace{N}_{\text{scan $\alpha$}}) \label{eq:O(n)}
\nonumber \\
& \sim &
O(N^{3-1/d} e^{\lambda(2d - 1)} + N^{4-3/d})
\label{eq:O(n)summary}
\end{eqnarray}
\narrowtext
(the second line summarizes the terms that dominate in different
limits). Understanding this expression gives us important information
about the practicality of Bogomolny's method.

First, the time needed to diagonalize the $T$ matrix does not
dominate when calculating highly excited states of two degrees of
freedom systems; this is contrasted to the case of matrix mechanics
where diagonalizing the Hamiltonian is virtually {\em all} of the
work. The reason is that the $T$ matrix is smaller than the
Hamiltonian; it operates on functions that have one dimension fewer
than the full quantum mechanical eigenstates so a smaller basis set
is adequate. However, for systems with more than two degrees of
freedom, diagonalizing $T$ is the dominant part of the work of the
algorithm; the advantage of smaller matrix size is overtaken by the
disadvantage that the matrix must be diagonalized $O(N)$ times.

Second, the effort of implementing the semiclassical method increases
with increasing chaos ($\lambda$), a fact that should be obvious
given that the method relies on classical trajectories. By contrast,
the dependence of the effort of a direct diagonalization of $H$ on
$\lambda$ is less explicit. As the degree of chaos is increased, it
typically becomes necessary to include more and more quantum
mechanical basis states in the matrix representation of the
Hamiltonian in order to get the same number of eigenenergies to
converge. Therefore, in practice matrix mechanics also becomes more
effort as the degree of chaos is increased. Nevertheless, a
quantitative estimate of the scaling would be tricky and will not be
attempted here.

Third, comparing expression (\ref{eq:O(n)summary}) against the matrix
mechanical result of $O(N^3)$, we see that Bogomolny's method should
be {\em faster} than matrix diagonalization at getting high-$N$
states when $d=2$, comparable when $d=3$, and poorer for $d\geq4$.

\subsection{Searching for eigen\protect\bbox{classicities} instead of
eigen-energies}
\label{sec:eigenclassicities}

So far we have been coy about specifying what we mean by the
parameters denoted by $\alpha$. In fact, $\alpha$ can represent any
external parameters that enter the Schr\"odinger equation---$E$,
$\hbar$, or parameters affecting the form of the Hamiltonian itself.
The key point is that the quantization condition (\ref{eq:invariant})
is not attainable for arbitrary parameters; it can only be satisfied
when there happens to be an eigenstate at that choice of parameters.
So $T_\alpha$ can only show the presence or absence of a quantum
eigenstate (by respectively having or not having an eigenvalue that
equals unity) at the {\em one particular} point in parameter space at
which it was computed. This is why we need to compute $T_\alpha$ many
times, for various selections of $\alpha$, in our search for
eigenstates of the system.

Normally one would vary only $E$, in which case unit eigenvalues of
$T_E$ mark eigen{\em energies} of the system and the usual energy
spectrum is produced. However, it should be clear that it is also
possible to fix $E$ and vary some other parameter of the problem. One
can even vary several of the parameters simultaneously.

In fact, if one varies several parameters simultaneously, while at
the same time keeping them in a carefully chosen relationship to one
another, one can arrange that the classical trajectories are left
unchanged (or maybe trivially rescaled) despite the change. {\em
Scalable potentials} (such as billiards) show a particularly simple
version of this effect---the classical trajectories scale trivially
as the energy itself is changed. If we find such a scaling
combination of parameters, we will only need to compute a mesh of
classical trajectories once, then reuse them as necessary to
calculate $T$ for many parameter values. Thus we would be able to
find {\em many eigenstates} (albeit not members of a single energy
spectrum) from {\em a single set} of classical trajectories. Having
to compute only a single set of trajectories, rather than a separate
set for each eigenstate to be found, significantly reduces the work
necessary to verify Bogomolny's method.

Treating Planck's constant as the variable parameter has the desired
effect. (Any reluctance to vary one of nature's fundamental constants
can be circumvented by noting that this operation is equivalent to
varying other parameters of the problem in synchrony. Details are
given in Appendix~\ref{sec:rescaling}.) Clearly Planck's constant has
no effect on the classical trajectories; one set of them can be
calculated and then used to calculate $T$ for any value of $\hbar$.

In fact, it is useful to think of $1/\hbar$ as the problem's {\em
classicity}. Increasing the classicity at constant energy shortens
the particle's de~Broglie wavelength; this, in turn, allows more
``nodes'' to fit on the energy shell, so that more highly excited,
more ``classical'' eigenstates result. In the sense that states of
higher classicity (everything else held constant) have higher
excitation numbers, classicity is analogous to energy, and it helps
to think of it as a kind of pseudoenergy---though beware that the
analogy is not exact (for example, eigenstates of classicity at fixed
energy are not orthogonal to one another).

In our numerical experiment outlined in Section~\ref{sec:Numerical},
we use this trick. We search for eigenstates of {\em fixed energy and
variable classicity}, producing an eigen{\em classicity} spectrum for
the system. In effect we are able to enjoy the computational leverage
that is usually associated with scalable potentials, but without
having to limit ourselves to a (non-generic) scalable potential.
Moreover, this trick allows us to change independently the two
parameters that are expected to affect the performance of the
semiclassical algorithm: $E$ (which sets the degree of chaos) and
$1/\hbar$ (which sets how close we are to the semiclassical limit).

\section{Computing Exact Eigenclassicities}
\label{sec:exact}

To evaluate the accuracy of Bogomolny's method for calculating
approximate eigenclassicities, it was necessary to compute the exact
eigenclassicities of our system for reference. This section contains
a brief description of that computation, preceded by some comments
about the eigenclassicity problem in general.

Solving the quantum eigenclassicity problem is a bit more complicated
(or at least more unfamiliar) than solving a quantum eigenenergy
problem. The latter is given by the familiar equation
\begin{equation}
H \vert\Psi_i\rangle = E_i \vert\Psi_i\rangle\;.
\label{eq:eigenenergy}
\end{equation}
$\vert\Psi\rangle$ represents a full, quantum mechanical
wavefunction, and is to be distinguished from $\vert\psi\rangle$,
which represents a surface of section wavefunction like the ones
operated on by the $T$ operator. After some complete set of basis
states $\{\vert n\rangle\}$ is chosen, and the matrix elements
$\langle n\vert H\vert n'\rangle$ are computed, the problem reduces
to an eigenvalue (matrix diagonalization) problem,
\[ \sum_{n'}
\langle n\vert H\vert n'\rangle
\langle n'\vert\Psi_i\rangle=
E_i\langle n\vert\Psi_i\rangle\;. \]

On the other hand, the eigenclassicity problem, even for a
Hamiltonian that is equal to kinetic energy plus potential energy,
takes a different form. One must rearrange Eq.~(\ref{eq:eigenenergy})
to isolate $1/\hbar$. To do this, we must ``look inside'' $H$:
\begin{eqnarray*}
H & = & \sum_i {1\over2} (\hbar k_i)^2 + V(\vec{q}) \\
& = & \hbar^2 K + V
\end{eqnarray*}
where $k_i$ are the wave numbers $p_i/\hbar$ associated with the
momenta, and $K= {1\over2} \sum k_i^2$ is a reduced kinetic energy in
terms of wave numbers. In terms of those quantities, the
eigenclassicity equation is
\begin{equation}
K \vert\Psi_i\rangle=
\left(\frac{1}{\hbar_i}\right)^2 (E-V) \vert\Psi_i\rangle\;.
\label{eq:eigenclassicity}
\end{equation}
Here $E$ is taken to be constant, and we look for the discrete values
of the classicity $1/\hbar_i$, and non-zero eigenvectors
$\vert\Psi_i\rangle$, for which the equation holds. Since the
wavevectors on both the left and the right sides of the equality are
multiplied by operators, this results in a generalized eigenvalue
problem. In some basis, the problem that needs to be solved is
\begin{equation}
\sum_{n'} \langle n\vert K\vert n'\rangle \langle
n'\vert\Psi_i\rangle=
\left(\frac{1}{\hbar_i}\right)^2
\sum_{n'} \langle n\vert (E-V)\vert n'\rangle \langle
n'\vert\Psi_i\rangle\;.
\label{eq:eigenclassicity_matrix}
\end{equation}

The eigenvectors corresponding to different eigenclassicities are not
orthogonal in the usual way; instead of satisfying
$\langle\Psi_i\vert \Psi_j\rangle= \delta_{ij}$, they satisfy
$\langle\Psi_i\vert K\vert\Psi_j\rangle= \delta_{ij}$ and
$\langle\Psi_i\vert (E-V)\vert\Psi_j\rangle= \delta_{ij}$. Also note
that the operator $(E-V)$ is not positive-definite, and in fact
$(1/\hbar)^2$ can possess negative solutions (though of course only
the positive solutions are physically meaningful).

\begin{figure}
\centerline{
\hbox{\psfig{file=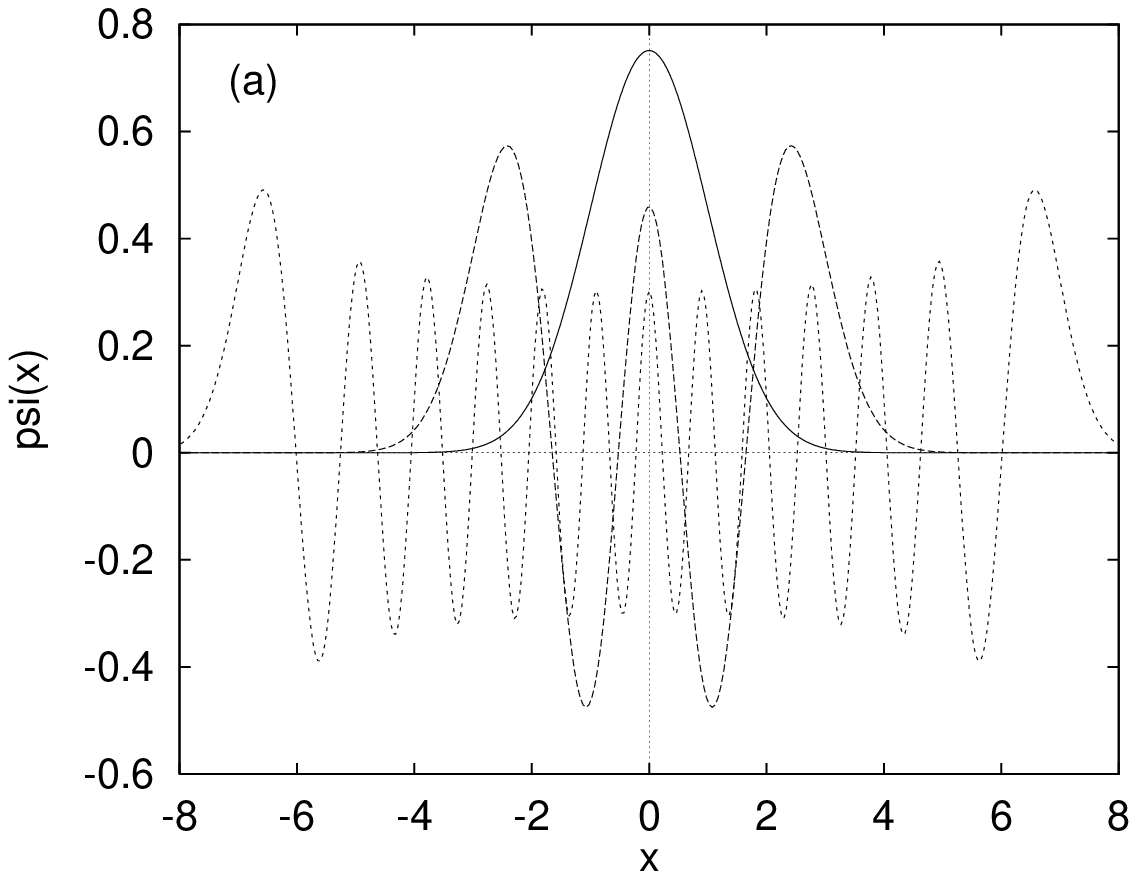,width=7.5cm}
\psfig{file=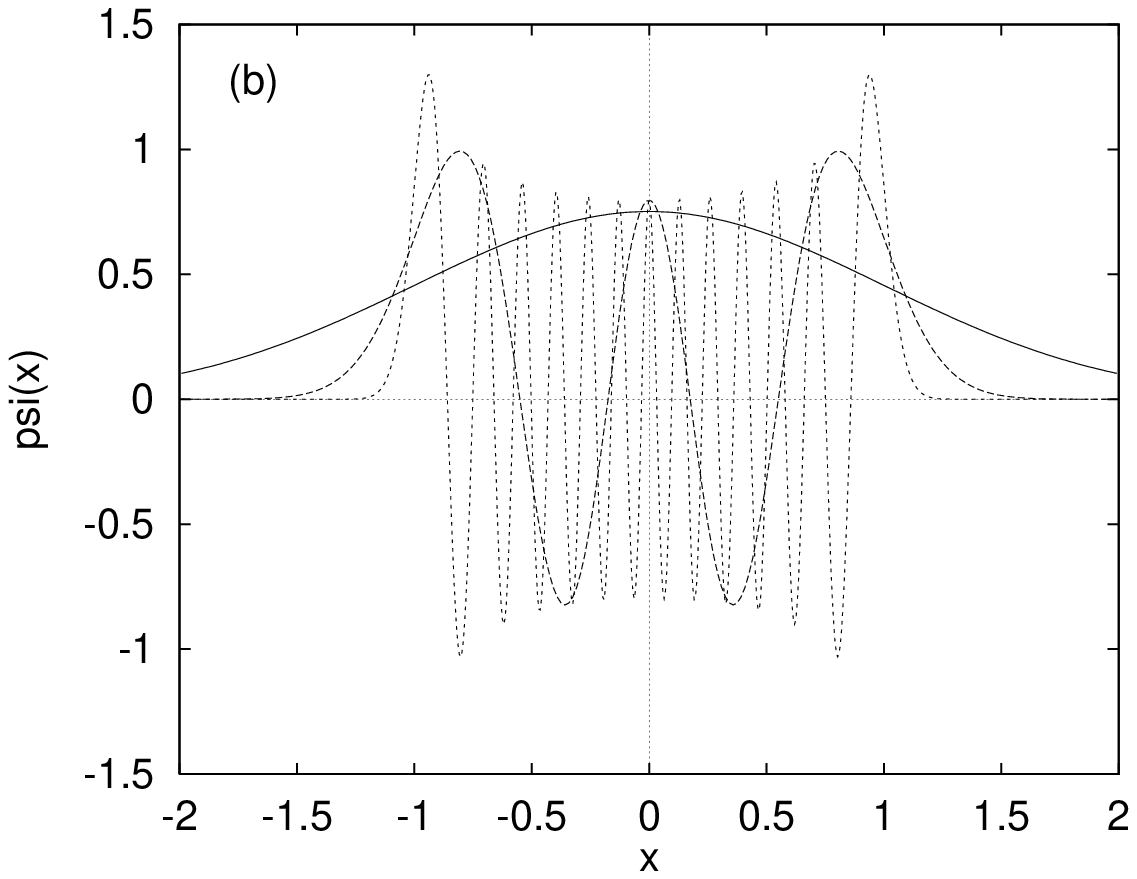,width=7.5cm}
}
}
\caption[Energy vs.\ classicity eigenfunctions for the 1-D
SHO]{Energy eigenfunctions vs.\ classicity eigenfunctions for the 1-D
harmonic oscillator $H= {1\over2} p^2+ {1\over2} x^2$. Each plot
shows the ground state and the 4th and 24th excited states.
(a)~Energy eigenfunctions (calculated at fixed $\hbar=1$) get wider
as the number of excitations increases, because the energetically
allowed region grows. (b)~Classicity eigenfunctions (calculated at
constant $E= \frac{1}{2}$), on the other hand, all have the same
classical turning points, and states of {\em lower} excitation number
turn out to be able to tunnel further into the classically forbidden
region.}
\label{fig:SHOeigenfunctions}
\end{figure}

The advantage of changing Planck's constant rather than the energy is
that the degree to which the system is quantum mechanical or
classical is a parameter that can be adjusted without changing the
classical dynamics of the system. At constant energy, states of
higher classicity are more highly excited in the sense of having
higher quantum numbers and more complicated nodal patterns. They do
not tunnel as well into classically forbidden regions of phase space.
They can be made more compact. In all of these senses, states of
higher ``classicity'' are more classical than states of lower
classicity---hence the choice of the name. To illustrate these
properties, Fig.~\ref{fig:SHOeigenfunctions} shows a few eigenenergy
states of the 1-dimensional harmonic oscillator, and contrasts them
to the analogous eigenclassicity states.

Since our semiclassical computation produced predictions of
eigenclassicities, we needed to compute exact quantum eiganclassicity
spectra of the Nelson$_2$ potential for comparison, in the manner
described above. We used a basis of 2-D harmonic oscillator
wavefunctions deformed to follow the parabola $y={1\over2} x^2$. (The
method is the same as that used in Ref.~\cite{Pro92}.) In this basis
the matrices of interest are banded. We optimized the horizontal and
vertical scale lengths of the basis functions until each
diagonalization could be done with a basis of a few thousand
functions. Finally, we checked the validity of the diagonalization by
comparing the resulting spectral staircase to the Thomas-Fermi
smoothed staircase, and using only states significantly below the
classicity at which those curves diverge. This procedure yields
eigenclassicities that are accurate to a small fraction of the mean
level spacing in the range of interest.

We will henceforth call the computed quantum mechanical values the
``exact'' values---not in reference to their numerical virtues, but
rather because they are computed on the basis of an {\em exact}
theory---as opposed to a semiclassical approximation.

\section{Numerical Experiment}
\label{sec:Numerical}

\subsection{The model system}
\label{sec:model}

We applied Bogomolny's method to the case of a particle moving in a
smooth nonlinear oscillator with Hamiltonian
\[ H = {1\over2} \left( p_x^2 + p_y^2 \right) + {1\over2}\omega^2 x^2
+ {1\over2} \left( y - {1\over2} x^2 \right)^2\;.
\]
Aside from a minor rescaling of variables discussed in
Appendix~\ref{sec:rescaling}, this system is identical to the
``Nelson'' potential studied by Baranger and Davies \cite{BarDav87}
and Provost \cite{Pro92}; to eliminate confusion we will refer to our
rescaled potential as ``Nelson$_2$.'' We fixed the value of
$\omega^2=0.05$ (the same value as used by those authors), and used
the $y$-axis as our surface of section.

\begin{figure}
\centerline{
\psfig{file=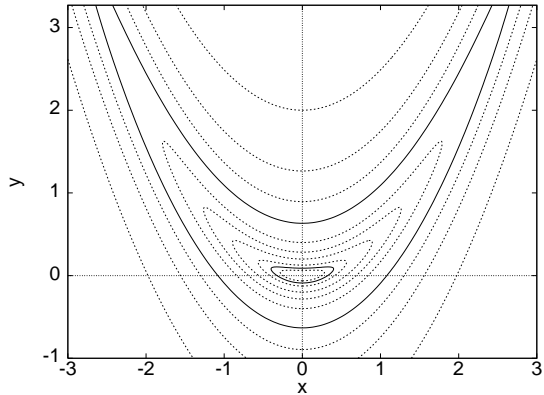,width=7.5cm}
}
\caption[Some contours of the Nelson$_2$ potential]{Some contours of
the Nelson$_2$ potential. The two solid contour lines are for the
energies $E=0.004$ and $E=0.2$, which were chosen for the numerical
experiments; the former is mostly regular, while the latter is mostly
chaotic. The surface of section used was the $y$-axis.}
\label{fig:contours}
\end{figure}

The system is an anisotropic harmonic oscillator elongated along the
$x$-direction which has been bent up along the parabola $y= {1\over2}
x^2$; some contour lines of this potential are shown in
Fig.~\ref{fig:contours}.
The system has a rich periodic orbit structure \cite{BarDav87} and is
bound at all energies. As energy is increased the particle explores
more and more of the curved ``horns'' and the motion becomes
increasingly chaotic. The Thomas-Fermi classical estimate of the
number of even and odd states for our potential, including
corrections up to ${\cal O}(\hbar^2)$, is given by \cite{Pro92}
\widetext
\begin{equation}
N(<E \text{ or } <(1/\hbar)) = \frac{E^2}{4 \hbar^2 \omega}
\left\{ 1 \pm \frac{\hbar\omega}{E}
- \frac{\hbar^2}{12}
\left[\frac{\omega^2 + 1}{E^2} + \frac{1}{\omega^2 E} \right]
\right\};
\label{eq:ThomasFermi}
\end{equation}
\narrowtext
the plus and minus correspond to the expressions for even and odd
parity states, respectively. To first order the number of states
increases quadratically with both energy and classicity.

We computed eigenclassicity spectra for two different values of
energy: $E=0.004$, where the system is predominantly regular; and
$E=0.2$, where it is predominantly chaotic. Parameters of that
computation are summarized in Table~\ref{tbl:classical_calcs}.
\begin{table}
\caption[Parameters for semiclassical computations]{
Parameters for semiclassical computations of eigenclassicities of
the Nelson$_2$ potential. In order to reproduce the $T$ operator
eigenvalue curves accurately, many times more $T$ diagonalizations
were done than would have been needed only to isolate the
eigenclassicities. Also, the number of trajectories used for the
regular regime computation was considerably higher than actually
needed.
}
\begin{tabular}{p{3.0in}rr}
Energy & 0.004 & 0.2 \\
Classicity range & 0--4000 & 0--80 \\
\raggedright
\# states in range (even and odd parity) & 574 & 572 \\
\raggedright
\# of classical trajectories used & $735\times735$ & $945\times945$
\\
\# of basis states used & 36 & 36 \\
\raggedright
\# of $T$ diagonalizations done & 3845 & 9188 \\
\end{tabular}
\label{tbl:classical_calcs}
\end{table}
For the two energies we calculated all of the eigenclassicities, of
both parities, in the ranges $0\leq (1/\hbar) \leq4000$ and $0\leq
(1/\hbar) \leq80$, respectively; a total of 574 and 572 states fall
in those ranges. Details are contained in the following sections.

\subsection{The semiclassical eigenclassicity spectrum}
\label{sec:SCspectrum}

In applying Bogomolny's method, we took advantage of the potential's
mirror symmetry about the $y$-axis by using the $T$ operator
associated with half-Poincar\'e mapping trajectories, as discussed in
Section~\ref{sec:symmetries}. Thus eigenstates of the system are
expected to occur at values of $1/\hbar$ for which $T$ has an
eigenvalue of $\pm1$. The set of half-trajectories that we used were
started on a rectangular mesh of initial conditions in the allowed
$(y,\theta)$ plane, with $735 \times 735$ trajectories for $E=0.004$,
and $945 \times 945$ trajectories for $E=0.2$.

We calculated $T$ as a matrix in a basis composed of simple harmonic
oscillator eigenfunctions on the $y$-axis (remember the basis need
only be complete on the surface of section). The length scale of the
basis functions was chosen such that they would be solutions to the
Schr\"odinger equation that would apply to motion on the $y$-axis,
\[ \left( - \frac{\hbar^2}{2} \frac{d^2}{dy^2}+
\frac{1}{2} y^2 \right) \phi_n(y)=
\hbar \left( n+ {1\over2}\right) \phi_n(y)\;, \]
with $\hbar$ chosen to correspond to the classicity used in that
particular $T$ calculation; thus, the basis states vary smoothly with
classicity. The number of basis states needed throughout was only 36.

To extract the curves of eigenvalues of $T$ as a function of
$1/\hbar$ which will be shown below, it is necessary to deduce which
eigenvalues at two adjacent values of classicity are connected on the
same eigenvalue curve. (This was the most tricky part of the
implementation of Bogomolny's method.) Our driver program
accomplishes this association by looking for unambiguous nearest
neighbors in the two sets; whenever there is ambiguity, the program
fills the gap by calculating a new set of eigenvalues at an
additional classicity between the first two. The process repeats
until all associations are unambiguous; thus the curves below are
reconstructed and plotted faithfully and in full detail.

\subsection{$T$ operator eigenvalues---qualitative observations}
\label{sec:eigenvalues}

\begin{figure}
\centerline{
\hbox{\psfig{file=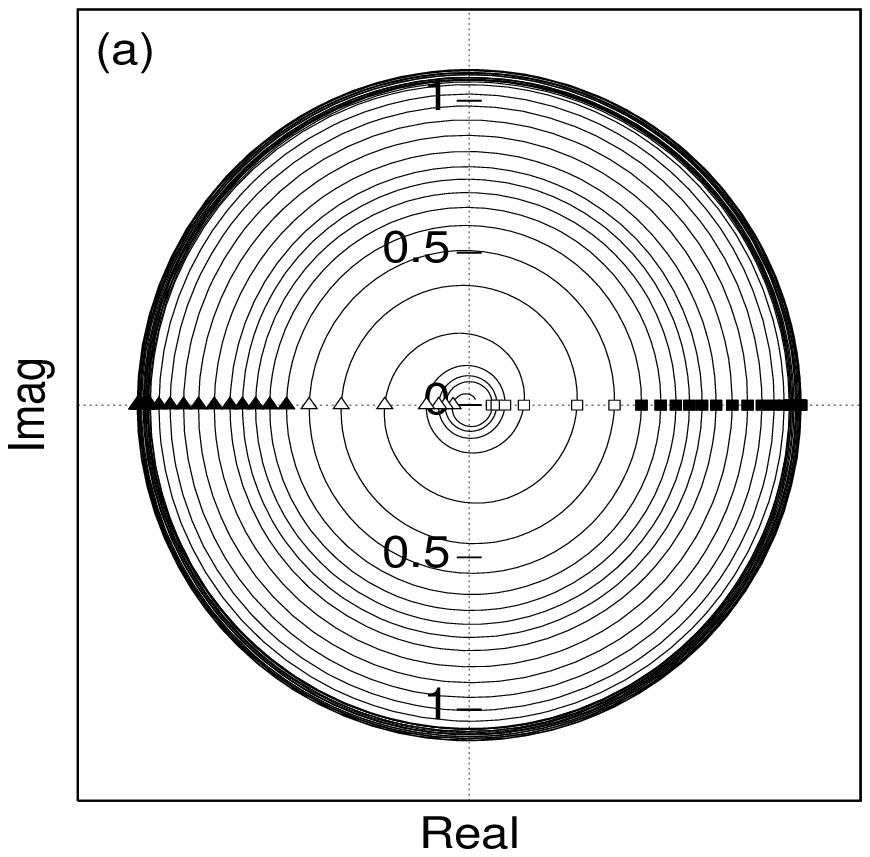,width=7.5cm}
\psfig{file=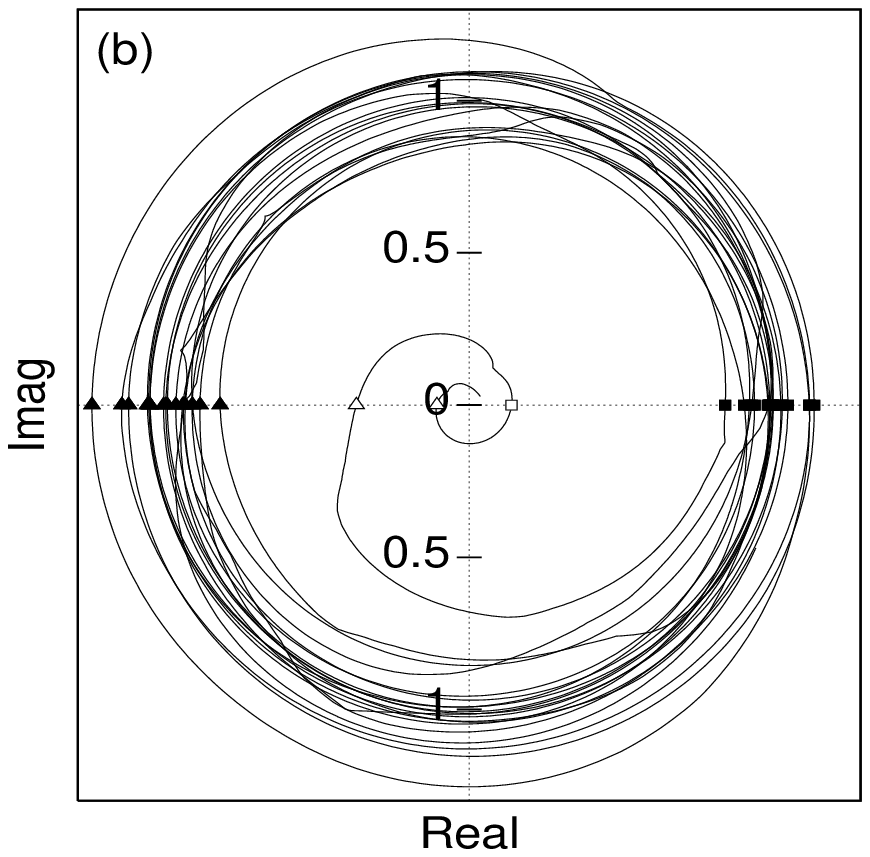,width=7.5cm}
}
}
\caption[Typical eigenvalue curves in the complex plane]{Typical
eigenvalue curves in the complex plane. Plotted are the curves that
two typical eigenvalues follow as the classicity ($1/\hbar$) is
scanned. The eigenvalues remain near the origin until the classicity
reaches a certain threshold (which is different for different
eigenvalues of $T$), at which time they begin to spiral out to the
unit circle. After that point, each time they cross the positive or
negative real axis, Bogomolny's theory predicts that the quantum
system should have an even or odd parity eigenstate, respectively.
Energies are: (a)~$E=0.004$; (b)~$E=0.2$. In each case, the fourth
$T$-operator eigenvalue to move from the origin to the unit circle is
plotted.}
\label{fig:complex3}
\end{figure}

\begin{figure}
\centerline{\psfig{file=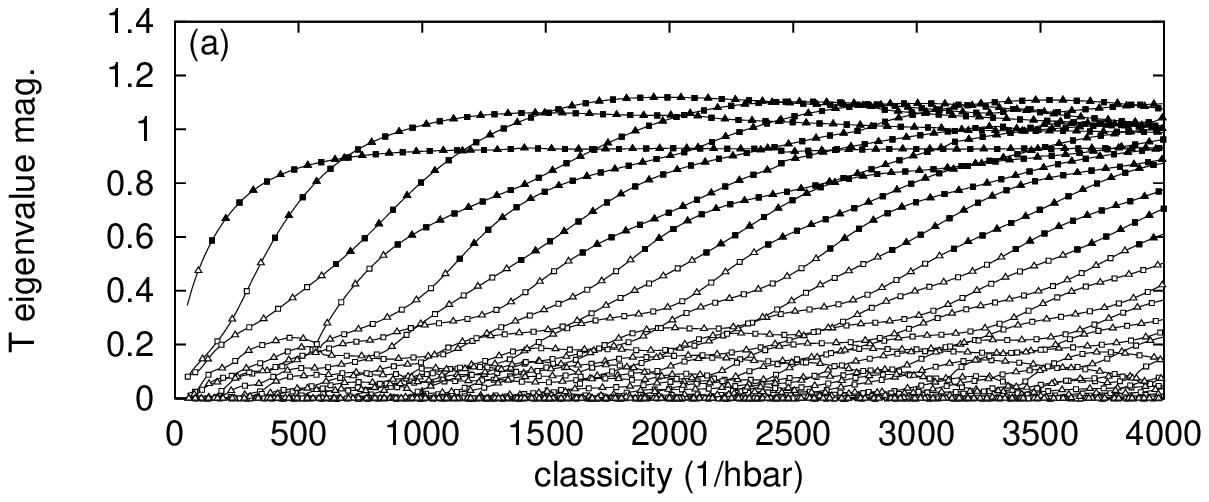,width=15.5cm}}
\centerline{\psfig{file=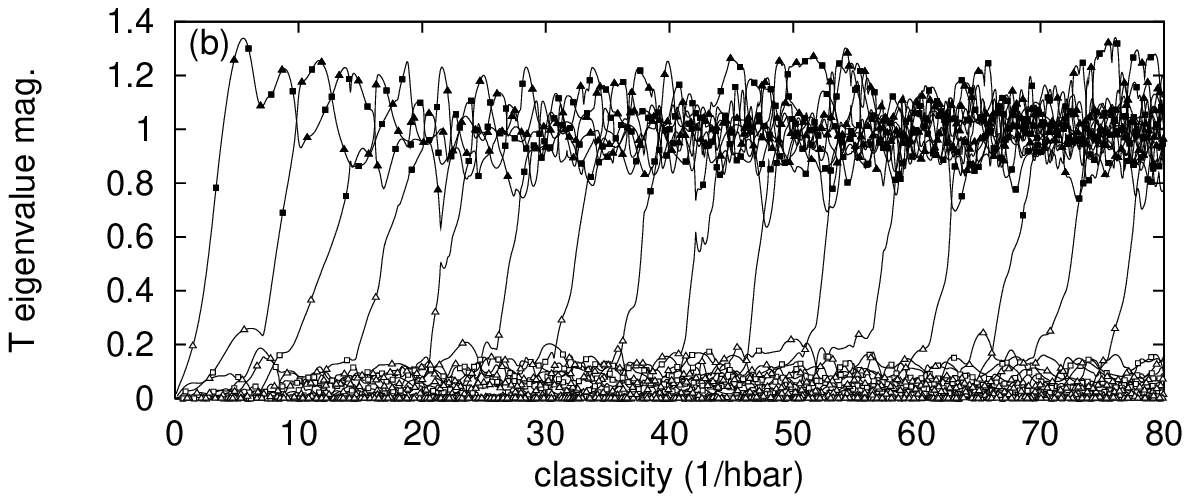,width=15.5cm}}
\caption[Magnitudes of the $T$-matrix eigenvalues as a function of
$1/\hbar$]{Magnitudes of the $T$-matrix eigenvalues as a function of
$1/\hbar$. As the classicity is increased, eigenvalues move, one by
one, from the origin towards the unit circle. Squares are plotted at
those points where the eigenvalues have phases of~$0$ (candidates to
be even parity eigenstates), triangles at phase~$\pi$ (odd
eigenstates). The symbols are plotted solid for those candidates
which turn out to be associated with true eigenstates of the quantum
system (``Class~1'' eigenvalues), and open for all others
(``Class~0'' eigenvalues). (a)~$E=0.004$, which is in the classically
regular regime. (b)~$E=0.2$, which is in the classically chaotic
regime. The horizontal axes are scaled such that the Thomas-Fermi
densities of states in the horizontal directions are equal for the
two energies. The plots predict the first 574 and 572 eigenstates,
respectively, of the full quantum mechanical system.}
\label{fig:mag}
\end{figure}

Much of the discussion of our numerical results concentrates on the
properties and behavior of the eigenvalues of the $T$ operator as a
function of classicity ($1/\hbar$). Recall that $T$ and its
eigenvalues can be computed for any value of classicity, so the
eigenvalues trace out continuous curves as the classicity is varied.
Figure~\ref{fig:complex3} shows examples of such curves in the
complex plane, for each of the two energies.
In order to see the behavior of the eigenvalues as the classicity is
varied, it is necessary to ``unroll'' the curves.
Figure~\ref{fig:mag} does this, showing the magnitudes of all of the
eigenvalues, as a function of classicity.

At any choice of parameters, $T$ has two relatively well-defined
classes of eigenvalues: those near the unit circle in the complex
plane and those near the origin. In the following discussion we
denote them as ``Class~1'' and ``Class~0'' eigenvalues, respectively.
The separation of eigenvalues into these classes turns out to be an
important feature of Bogomolny's method, and will be discussed in
detail below.

The general behavior of the eigenvalues is as follows: at very low
classicity, all of the eigenvalues are located in a cloud near the
origin (in Class~0). As the classicity is increased (corresponding to
higher excitations of the quantum system), they spiral out, one by
one, from the origin to an annulus near the unit circle (in Class~1).
Thereafter, they remain near the unit circle, continuing to rotate
counterclockwise. The behavior of the eigenvalues can be seen in
Fig.~\ref{fig:complex3}.

\subsubsection{The unitarity and dimension of $T$}

\begin{figure}
\centerline{
\hbox{\psfig{file=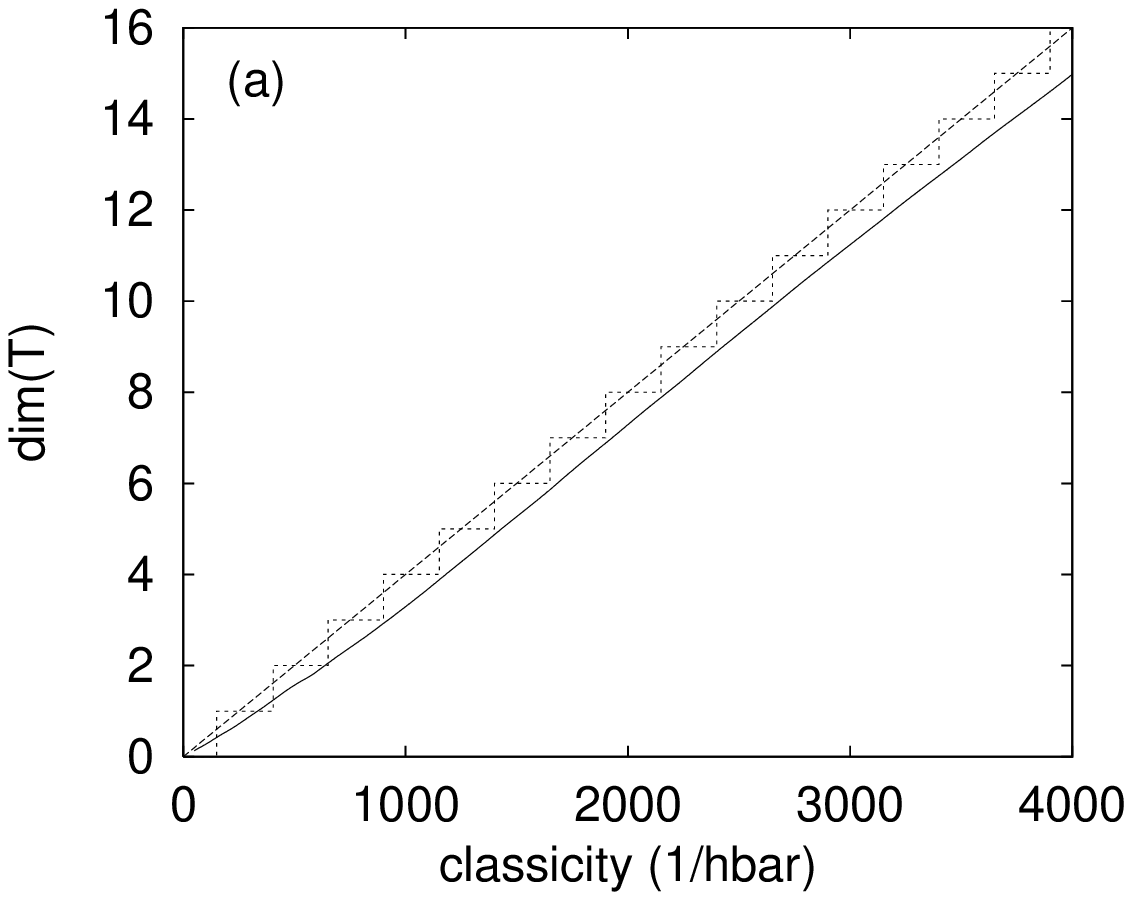,width=7.5cm}
\psfig{file=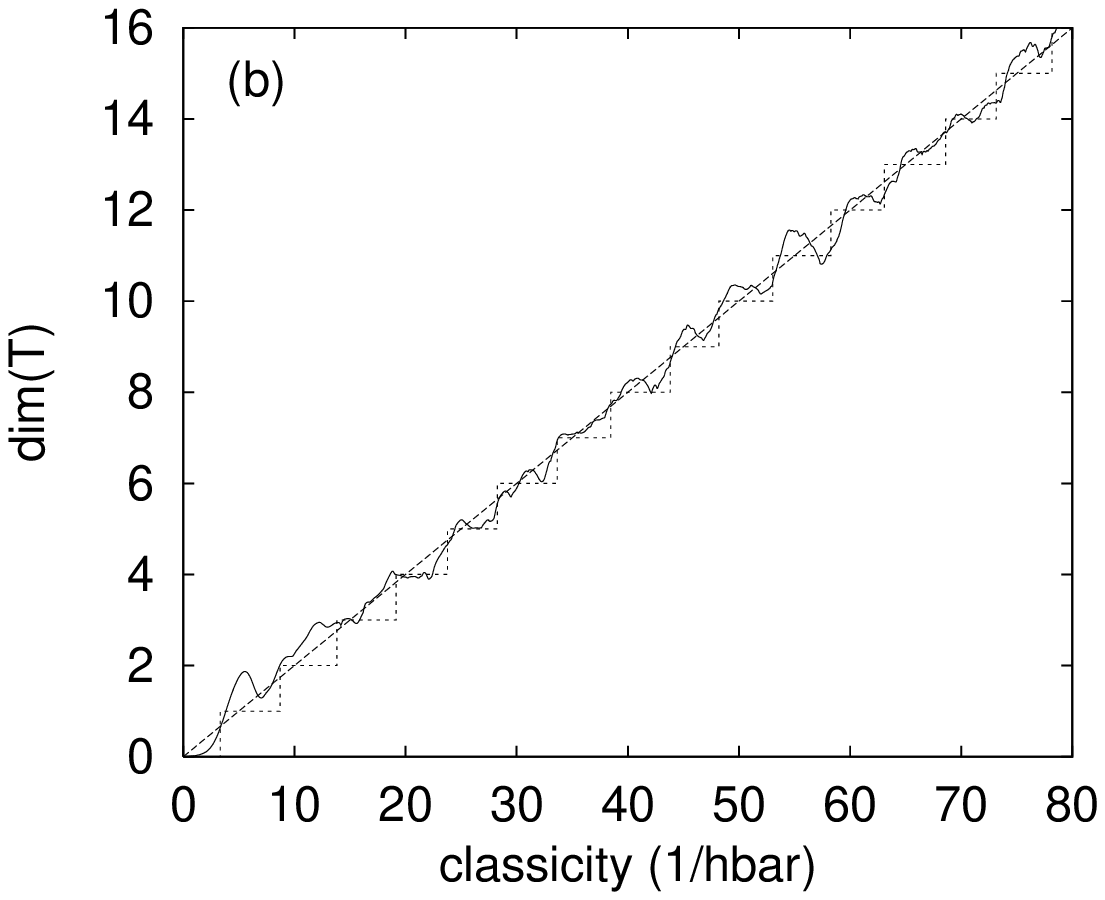,width=7.5cm}
}
}
\caption[Dimension of the $T$ operator as a function of
classicity]{Two notions of the dimension of the unitary part of the
desymmetrized $T$ operator are plotted as a function of classicity,
for (a)~$E=0.004$, (b)~$E=0.2$. The staircase is a count of how many
of the eigenvalues have predicted a quantum eigenstate up to that
point (roughly, the number of eigenvalues near the unit circle). The
solid curve shows ${\rm Tr}\, T^\dagger T$, which is a continuous
measure of the dimension of $T$. The dashed line shows Bogomolny's
theoretical prediction (Eq.~(\protect\ref{eq:Nelsondim})).}
\label{fig:Tdim}
\end{figure}

Bogomolny shows that, in the limit $\hbar \rightarrow 0$ and for $y$
and $y'$ on the classically allowed part of the surface of section,
$T$ is unitary in the sense that $\langle y'\vert T^\dagger T\vert
y\rangle= \delta(y'-y)$. The dimension of the ``unitary part'' of
$T$, however, is finite, and is given by Eq.~(\ref{eq:Tdimension}).
When Eq.~(\ref{eq:Tdimension}) is evaluated for the Nelson$_2$
potential, it gives the prediction that
\begin{equation}
\dim T = E \cdot (1/\hbar)\;.
\label{eq:Nelsondim}
\end{equation}
To the extent that the eigenvalues separate into classes as mentioned
above, then, the number of Class~1 eigenvalues at any given value of
parameters is given by Eq.~(\ref{eq:Nelsondim}). This prediction is
tested in Fig.~\ref{fig:Tdim}---the staircase functions are a count
of the number of eigenvalues that have migrated to Class~1 (more
precisely, what is plotted is the number of eigenvalues that have
predicted a quantum eigenstate).

This doesn't mean, however, that the $T$ matrix needs to be computed
in a basis of this size, which would require a manual truncation.
Rather, $T$ can be computed in a basis of arbitrarily large size, and
the number of Class~1 eigenvalues will obey Eq.~(\ref{eq:Nelsondim})
automatically. The rest of the eigenvalues---an infinite number of
them---remain in a cloud near the origin in Class~0, where they do
not affect the subsequent predictions. Correspondingly, a different,
continuous measure of the dimension of $T$ can be defined which does
not require a truncation or a counting of Class~1 eigenvalues:
\begin{equation}
\dim T \equiv {\rm Tr}\, T^\dagger T\;.
\label{eq:Tdim}
\end{equation}
This expression can be applied to computed $T$ matrices, and it
reflects the fact that the dimension varies continuously with
classicity. This measure is also plotted in Fig.~\ref{fig:Tdim}. As
can be seen, the agreement is quite good, even for low classicities.
It is also noteworthy that the dimension varies even more smoothly
than the curves of the individual eigenvalue magnitudes
(Fig.~\ref{fig:mag}); this is because variations in the magnitude of
one eigenvalue tend to be negatively correlated with variations in
those of another.

\subsubsection{Finding the quantum spectrum from $T$ eigenvalues}

Recall that the criterion for an even eigenstate of the system is
that $T$ have an eigenvalue equal to $1$ (for brevity we temporarily
ignore the odd-parity eigenstates that occur at eigenvalues of $-1$).
Of course, Bogomolny's method is a semiclassical approximation, which
always entails the use of a stationary phase approximation to the
exact Feynman path integral. Thus we shouldn't be surprised that
although the eigenvalues come near $1$, they never exactly equal $1$.
Therefore, a more robust criterion is needed than ``equality to 1.''

What is clear is that quantum eigenstates should only be associated
with Class~1 eigenvalues---the eigenvalues that have magnitudes of
approximately unity. As the classicity is increased, these
eigenvalues rotate counterclockwise along the unit circle. On each
rotation they pass close to 1 and ``generate'' an eigenstate. Leaving
for later the subtleties of determining exactly which eigenvalues are
in Class~1 and which are Class~0, we still need to decide at exactly
which point during the rotation of a Class~1 eigenvalue the
eigenstate is predicted to occur. At least three possible criteria
suggest themselves:
\begin{enumerate}
\item the point at which $\det(1-T)=0$ is most nearly fulfilled;
\item the point where a Class~1 eigenvalue closest approaches $1$;
and
\item the point where a Class~1 eigenvalue crosses the positive real
axis (has a phase of $0$).
\end{enumerate}

\begin{figure}
\centerline{
\hbox{\psfig{file=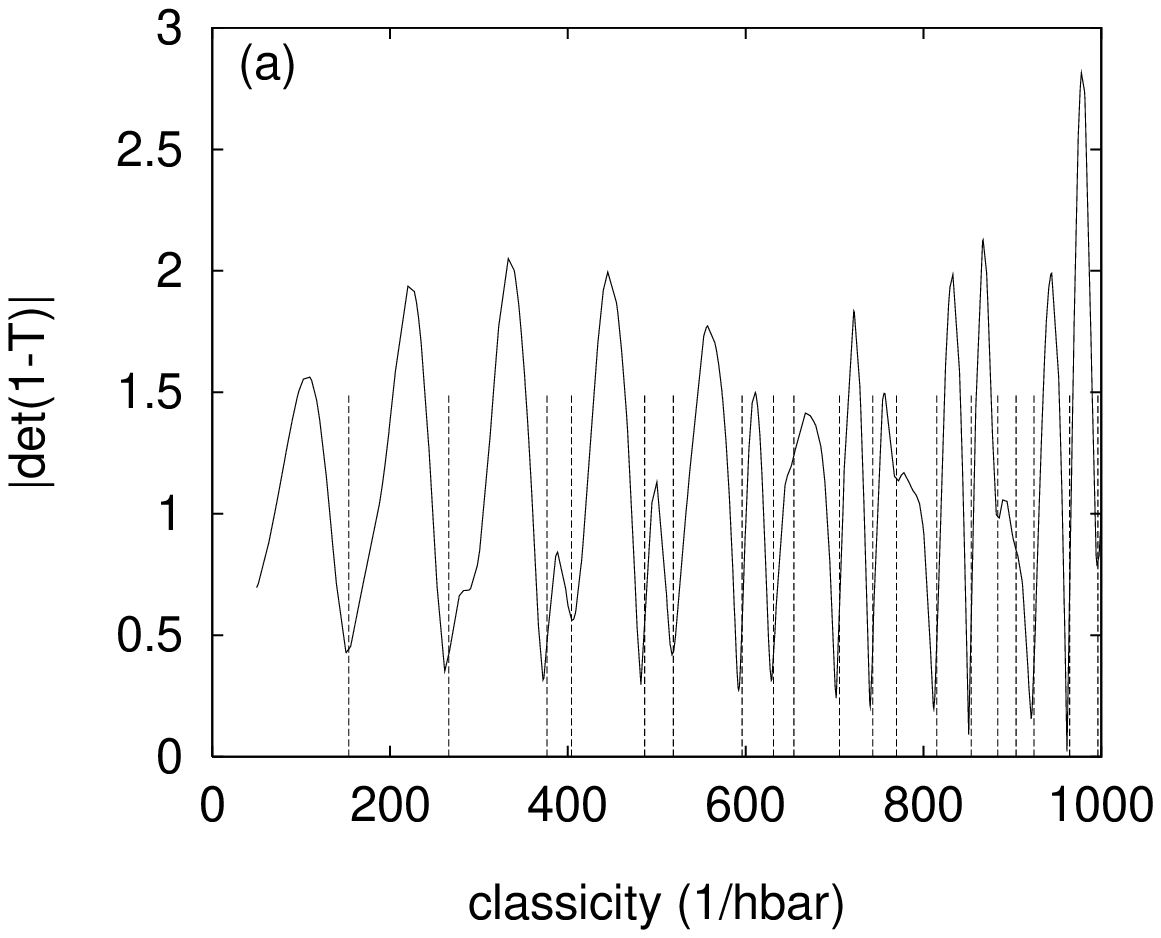,width=7.5cm}
\psfig{file=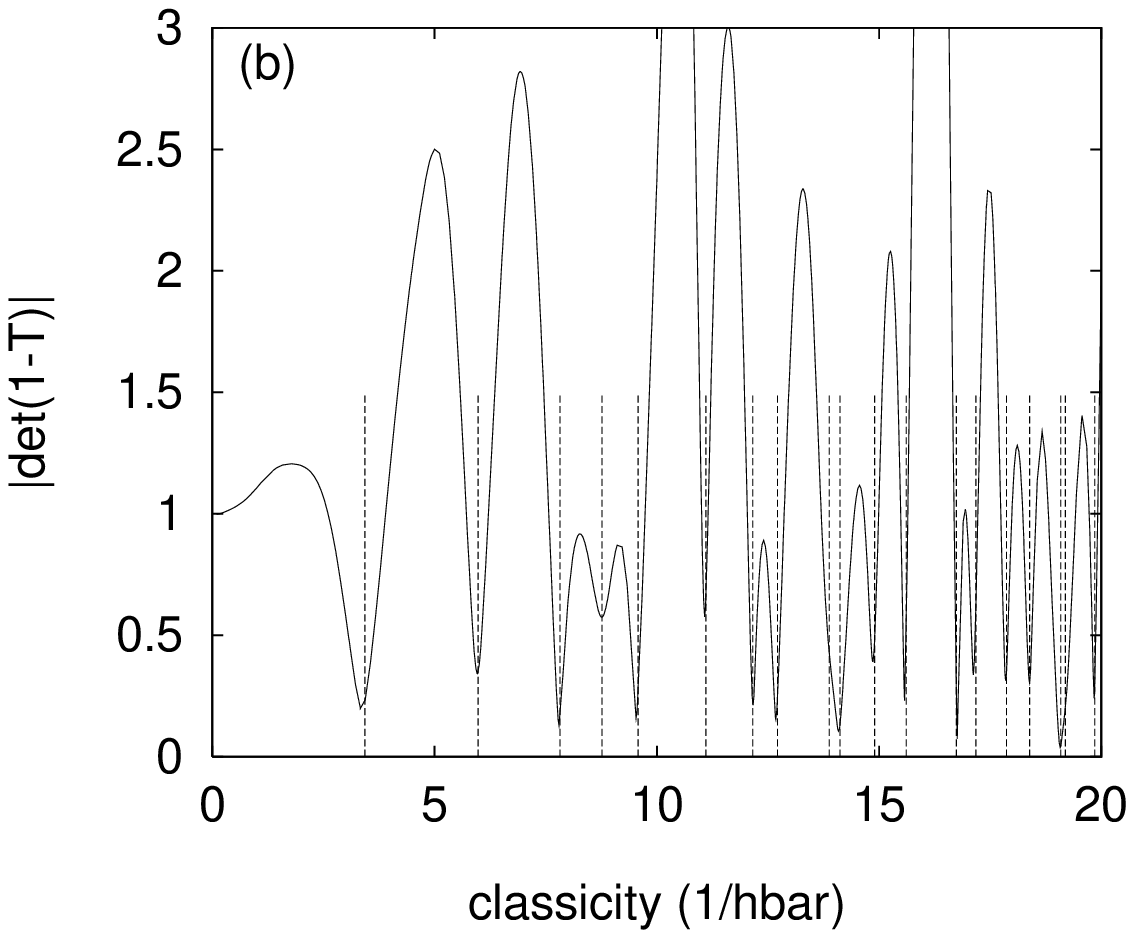,width=7.5cm}
}
}
\centerline{
\hbox{\psfig{file=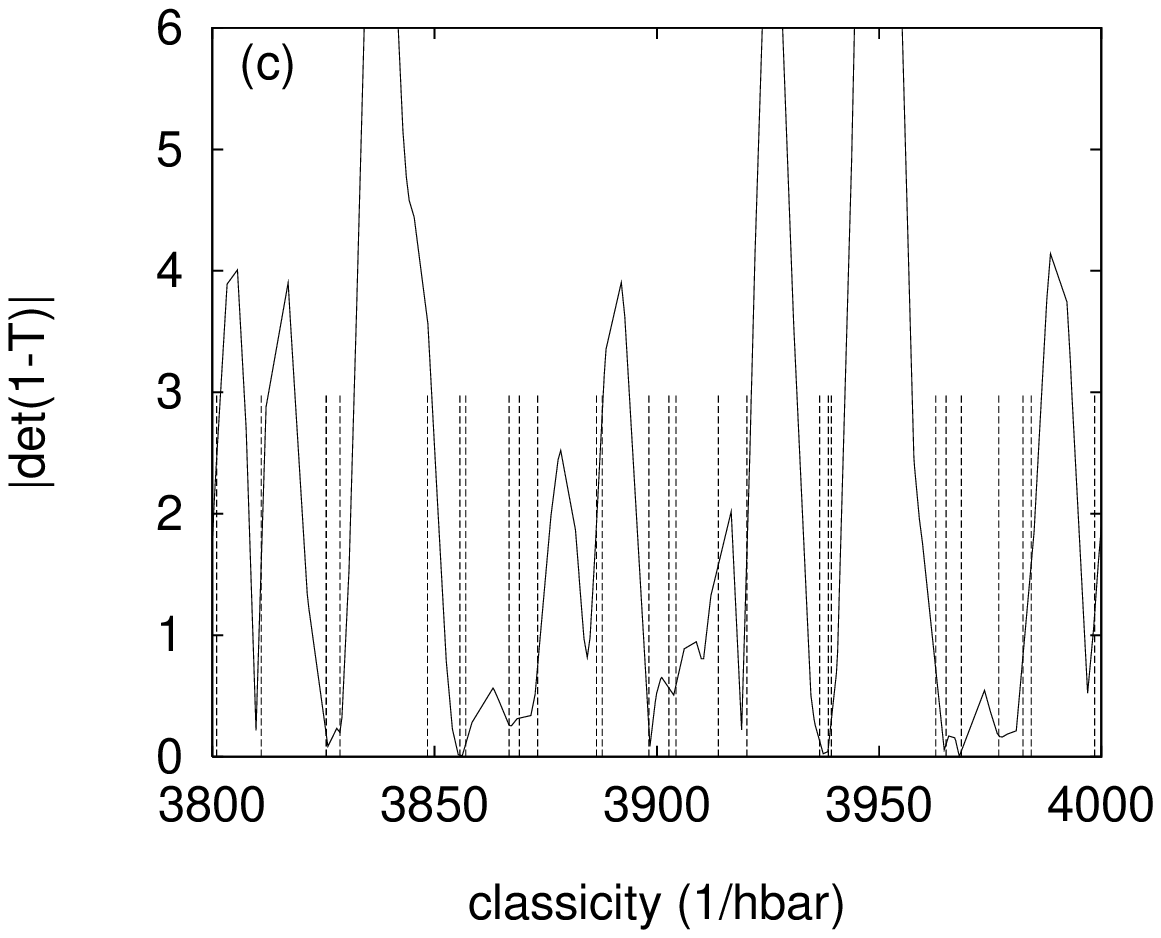,width=7.5cm}
\psfig{file=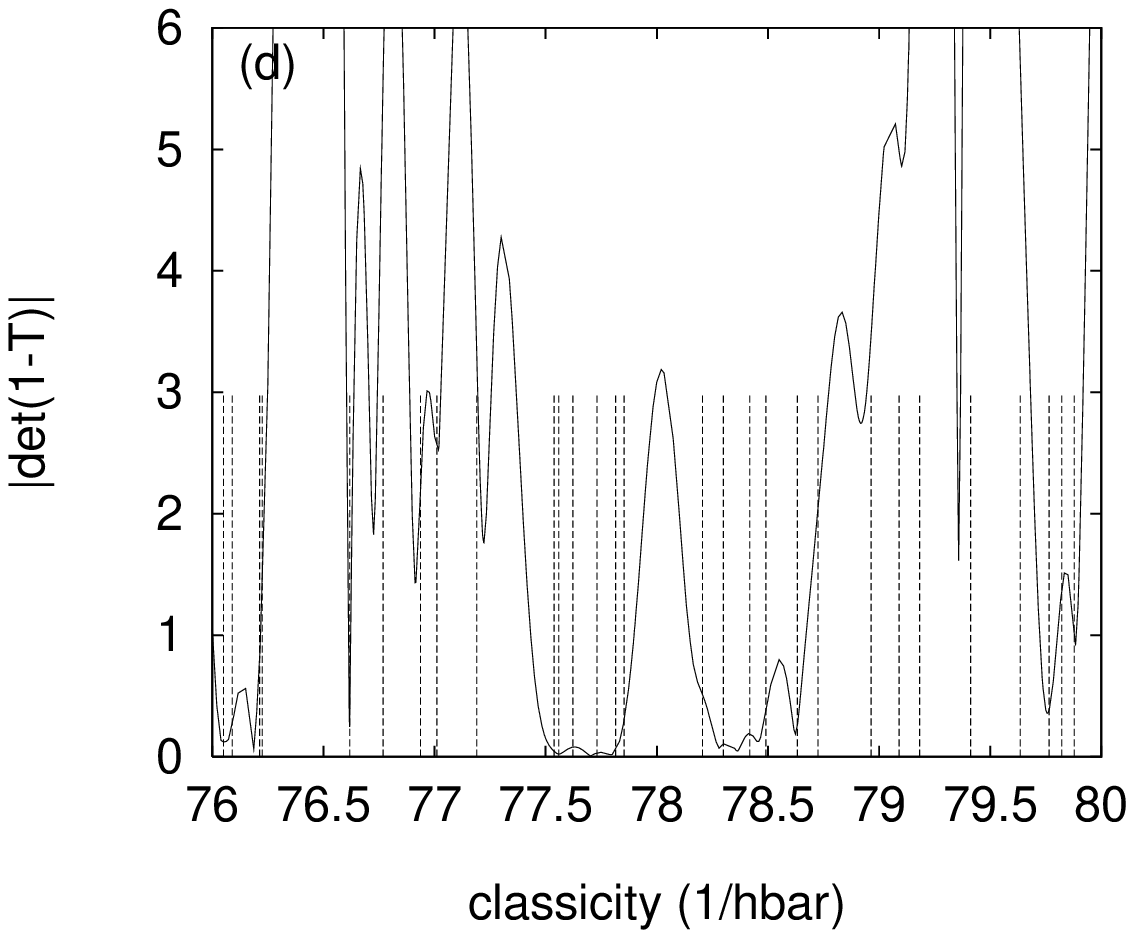,width=7.5cm}
}
}
\caption[$\vert\det(1-T)\vert$ versus
classicity]{$\vert\det(1-T)\vert$ versus classicity. The determinant
of $(1-T)$ is supposed to equal zero whenever the quantum mechanical
system has an even eigenstate. Since the determinant is complex, we
plot its magnitude as a function of classicity (the solid line) and
indicate the exact eigenvalues with vertical dashed lines. Shown are:
the lowest 19 even eigenstates for (a)~$E=0.004$, (b)~$E=0.2$; and
about 25 eigenstates around the 275th even eigenstate for
(c)~$E=0.004$, (d)~$E=0.2$. Although the minima match up reasonably
well for the low-lying states, there is clearly little predictive
power in the semiclassical determinant for highly-excited states. We
shall see that monitoring individual eigenvalues of $T$ works
better.}
\label{fig:Tdet}
\end{figure}

Although criterion~1 is the one emphasized by Bogomolny, it turns out
to be unsatisfactory. The determinant mixes together information
about all of the eigenvalues of $T$, whereas eigenstates are each
associated with a single eigenvalue of $T$. Therefore, the minima
that are supposed to indicate eigenstates are sometimes obscured,
sometimes overlapping (and therefore indistinguishable), and often
not very close to zero. Figure~\ref{fig:Tdet} shows all of these
effects.
For the low-lying states, the minima do correlate reasonably well
with the eigenstates of the quantum system, though there are already
a few exceptions. However, for the highly excited states (here around
$n=275$), the association is dramatically degraded. {\em There is
clearly little hope of making unambiguous predictions (let alone
accurate ones) of highly-excited eigenstates of the quantum system
based on a plot of $\vert\det(1-T)\vert$.}

Criteria~2 and 3 produce results that differ only very slightly from
one another. Criterion~3 is more robust than 2 and, we believe, more
appropriate; therefore in our search for even parity eigenstates, we
concentrate on those points where the $T$ operator has a Class~1
eigenvalue with $\text{\em phase}=0$. (Odd eigenstates are similarly
found where a Class~1 eigenvalue has $\text{\em phase}=\pi$.)
Accordingly, in Fig.~\ref{fig:mag} we have placed symbols on the
curves whenever the eigenvalue crosses the real axis: squares and
triangles mark the points where the eigenvalues have phases of $0$ or
$\pi$, and which are thus candidates to be even or odd eigenstates,
respectively.

It is important to emphasize that there is {\em no ambiguity at all}
in the recipe for finding places where an eigenvalue crosses the real
axis. Through the course of our numerical explorations, we found that
the eigenvalues' phases increase monotonically as the classicity was
increased, except for a very few, brief exceptions. Since $T$ can be
computed and diagonalized at any value of classicity, this means that
any simple scheme suitable for finding a bracketed zero of a
continuous function suffices to pinpoint the classicity at which a
$T$ operator eigenvalue crosses the real axis of the complex plane.

Having now located all of the $\text{phase}=0$ points, it remains
only to be determined exactly which of those candidates correspond to
true quantum states. Equivalently, what we need is a criterion for
sorting eigenvalues into Class~1 and Class~0. How close to $1$ do the
eigenvalues need to be?

To investigate this question, we used our knowledge of the exact
spectrum to determine which of the candidates matched true states.
Thus {\em a posteriori} we were able to classify candidates
accurately as Class~1 (corresponding to a true quantum state) or
Class~0 (not associated with a quantum state). (In
Fig.~\ref{fig:mag}, the associated symbols are plotted solid if they
are in Class~1, open if Class~0.) But more importantly, we would like
to use our experience to describe how Class~1 eigenvalues could be
distinguished from Class~0 eigenvalues {\em a priori} in a purely
semiclassical calculation, without the benefit of knowing the exact
answers.

\begin{figure}
\centerline{
\hbox{\psfig{file=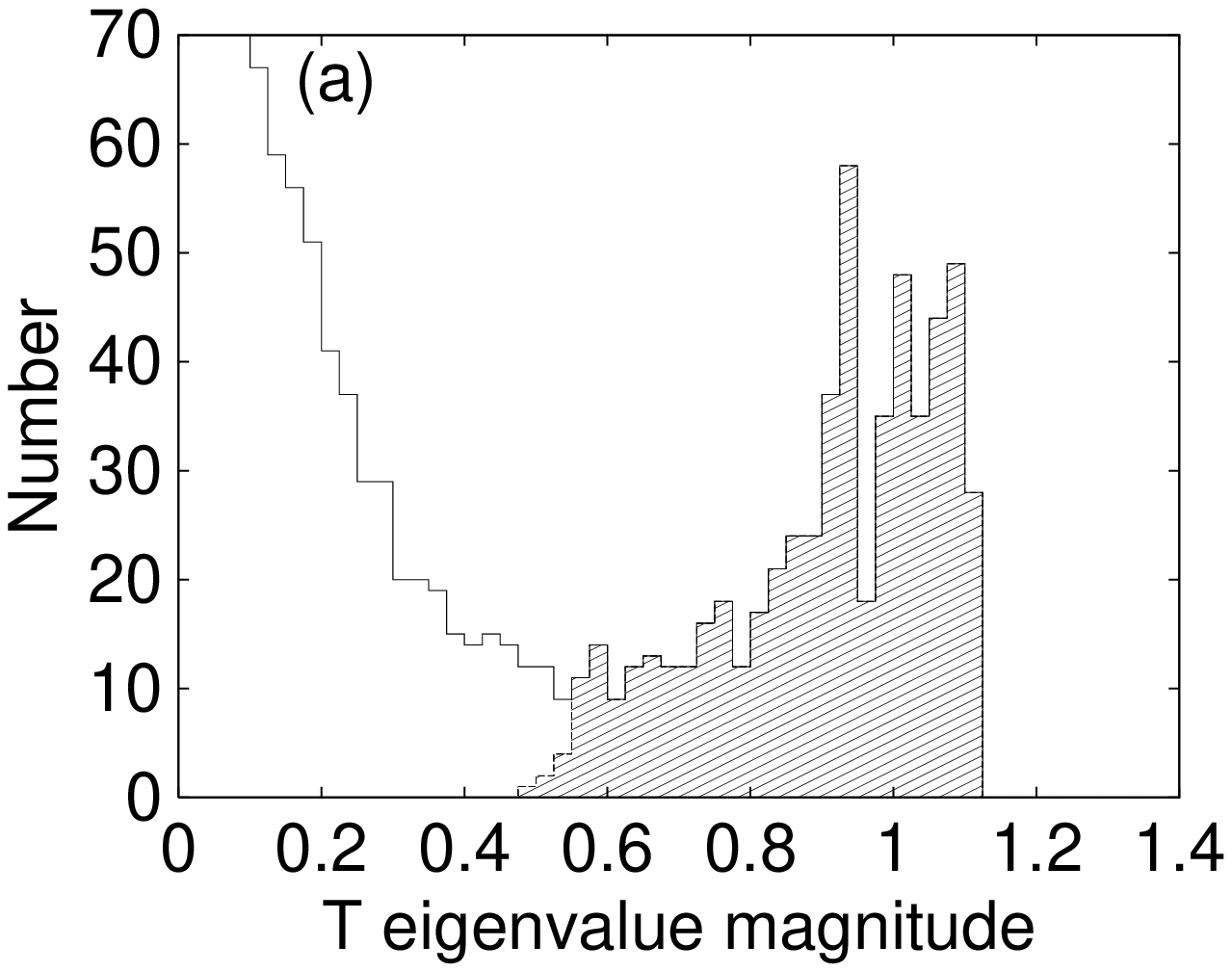,width=7.5cm}
\psfig{file=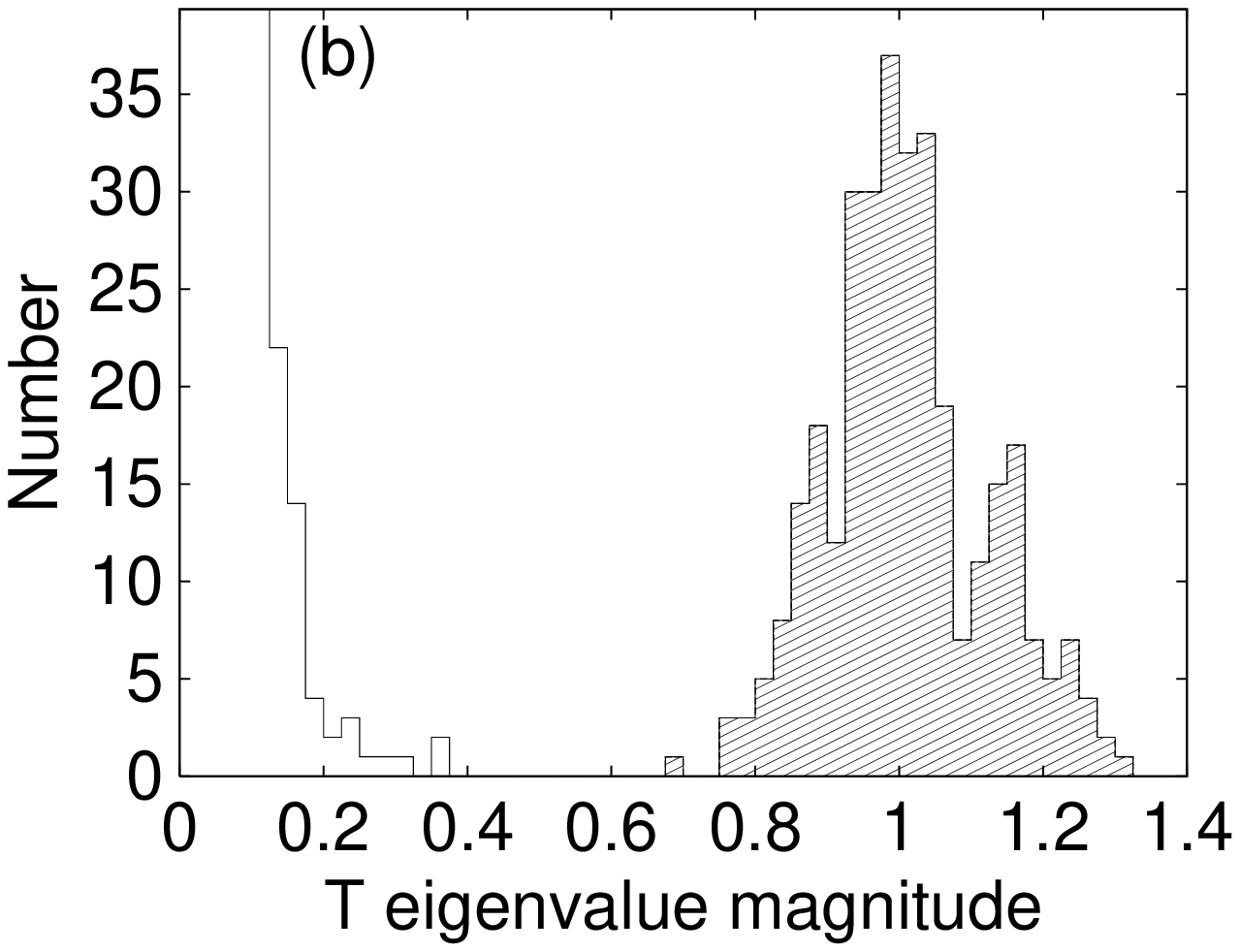,width=7.5cm}
}
}
\caption[The separation of $T$ eigenvalues into Class~0 and
Class~1]{The separation of eigenvalues into Class~0 and Class~1.
Plotted is the distribution of magnitudes of $T$ operator eigenvalues
which cross the (positive or negative) real axis. The shaded part of
the histogram contains Class~1 crossings, which are associated with
true quantum eigenstates; the unshaded part, Class~0 crossings.
(a)~$E=0.004$. In this (regular) regime, there is a transition region
around magnitude $0.5$, in which Class~0 and Class~1 eigenstates are
both present and thus difficult to distinguish by magnitude alone.
(b)~$E=0.2$. In this (chaotic) regime, the two classes are well
separated and Class~1 eigenvalues can easily be identified by their
larger magnitudes. Note that in both cases, there are an infinite
number of Class~0 crossings near zero magnitude.}
\label{fig:histogram}
\end{figure}

The first criterion is, of course, the magnitude of the eigenvalue
when it crosses the real axis. In the semiclassical limit, Class~1
eigenvalues are all supposed to have magnitude 1, and Class~0
eigenvalues, magnitude 0. At the finite classicity of our
experiments, there is considerable deviation from that ideal, as
shown in Fig.~\ref{fig:histogram}.
While it is true that the eigenvalues tend to cluster around either
the origin or the unit circle, the bands are pretty wide. Does the
width of the bands cause practical problems?

In the regular regime, it sometimes does: there is a range of
magnitudes, around 0.475--0.55, in which both Class~1 and Class~0
eigenvalues occur. In this overlap area, magnitude information is not
sufficient to classify the eigenvalues. Fortunately, only a few
percent of the eigenvalues fall into this uncertain range; the rest
are predicted unambiguously by Bogomolny's method. Even {\em in} this
range, it is likely that in many cases one could tell which of the
ambiguous eigenvalues need to be included in Class~1 by looking for
deficits in the semiclassical staircase as compared to the
Thomas-Fermi smoothed spectral staircase function, or deviations of
the staircase dimension of $T$ (as in Fig.~\ref{fig:Tdim}) from the
semiclassically expected result.

\begin{figure}
\centerline{
\psfig{file=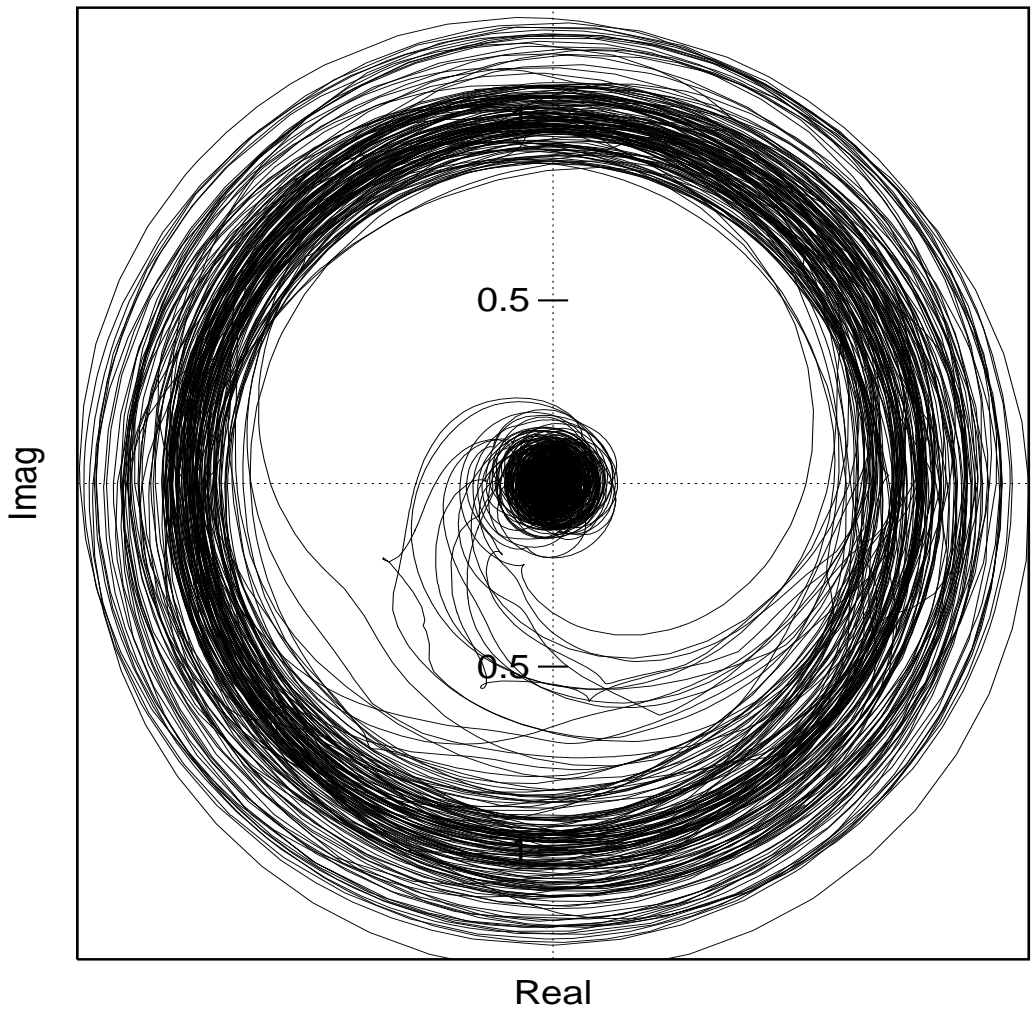,width=14cm,height=14cm}
}
\caption[$T$-matrix eigenvalues in the complex plane, for
$E=0.2$]{$T$-matrix eigenvalues in the complex plane, for $E=0.2$.
This figure is similar to Fig.~\protect\ref{fig:complex3}, except
that all of the eigenvalues of $T$ are shown on the same plot. The
striking feature revealed is that all of the eigenvalues spiral from
the origin to the unit circle along a single band in the lower half
of the complex plane. Within that rotation of $\pi$ radians, they
manage to make it all the way from Class~0 to Class~1; during their
brief transition, they are away from the real axis and so do not
produce crossings of uncertain class.}
\label{fig:complexh02}
\end{figure}

In the chaotic regime, the width of the bands causes no problem: it
is still easy to distinguish Class~1 from Class~0 eigenvalues,
because of the large gap separating them. In our experiment in the
chaotic regime, no Class~0 eigenvalues had magnitudes above 0.4, and
no Class~1 eigenvalues had magnitudes below 0.65. This fortuitous
circumstance is not only the result of the eigenvalues' spiraling
quickly from the origin to the unit circle; even a quick motion, if
it occurred near a real-axis crossing, would cause trouble. The other
important, and more surprising, property is that the eigenvalues all
spiral out along a relatively narrow band in the lower complex half
plane, and the entire journey is completed in less than the time it
takes for half a rotation around the origin (see
Fig.~\ref{fig:complexh02}).
As a result, there is {\em no ambiguity whatsoever} about the
semiclassical method's predictions for the location of
eigenclassicities in the chaotic regime. This is so important and
unexpected that it deserves to be emphasized: in the chaotic regime,
Bogomolny's method plus our criterion yield predictions of {\em every
single line} in the quantum spectrum, with {\em no spurious
predictions} whatsoever. This reliability is in contrast to that of
other semiclassical methods, which frequently fail to resolve
adjacent eigenstates and thereby leave doubt even about the number of
eigenstates in a spectrum.

Altogether, the first 574 eigenstates in the regular regime, and the
first 572 eigenstates in the chaotic regime (in each case, some even,
some odd parity) are reproduced by the data. The {\em accuracy} of
the semiclassical predictions will be discussed in
Section~\ref{sec:accuracy}, after a few more qualitative
observations.

\subsubsection{Quantum numbers from the semiclassical data}
\label{sec:qnumbers}

Each Class~1 eigenvalue of the $T$ operator produces many
eigenstates, one each time it rotates through the real axis.
Consequently it is possible to separate the eigenstates into groups,
based on which $T$ operator eigenvalue each one is associated with.
In Fig.~\ref{fig:mag}, eigenstates on a particular eigenvalue curve
thus can be considered to be members of the same group. Although the
groups are well defined in both regimes, they are truly meaningful
only in the regular regime.

For the mostly regular energy $E=0.004$, the system is nearly an
anisotropic harmonic oscillator, so it is nearly separable into $x$-
and $y$-motions. Eigenstates of the system can therefore be labeled
by two ``almost good'' quantum numbers, $n'_x$ and $n'_y$, which
count the number of excitations along and perpendicular to the
surface of section, respectively. The near separability is the reason
that the eigenvalue curves in the regular regime are so smooth and
unkinked; and the quantum numbers can be read off the picture as
well. All of the eigenstates on the first eigenvalue curve have
$n'_y=0$---that is, they have no excitations in the vertical
direction; those on the second curve have $n'_y=1$; on the third
curve, $n'_y=2$; etc. Meanwhile $n'_x$ can also be read off the
diagram: the first eigenstate on a particular curve has $n'_x=0$; the
second, $n'_x=1$; etc.

For the mostly chaotic energy $E=0.2$, however, the system is far
from separable, and it has no set of good quantum numbers. The
eigenstates still lie on continuous curves, but now the curves are
kinked and bent whenever two eigenvalues approach each other in the
complex plane. We see evidence that each interaction of two curves is
accompanied by an intermixing of the eigenstates' properties in the
same manner as happens at ``avoided crossings'' of energy levels,
seen when a quantum mechanical system's external parameter is scanned
adiabatically. So although the eigenstates are still connected by
eigenvalue curves, the eigenstates lying on a single curve do not
necessarily have similar properties, and the grouping by curves is
not helpful.

\subsection{Accuracy of eigenclassicity spectra}
\label{sec:accuracy}

We have now outlined all of the steps necessary to compute the
eigenclassicity spectrum predicted by Bogomolny's quantum surface of
section method. In order to check its accuracy, it was necessary to
decide, for each of the semiclassically computed eigenclassicities,
which of the exact eigenclassicities it was ``trying to predict.''
This we did manually by comparing the two spectra; usually the
eigenstates lined up so well that the correlation was obvious. When
two states were very close to one another, the further step of
comparing the exact and semiclassical surface of section
wavefunctions was taken; this almost always made it obvious how to
match up the numbers.

\begin{figure}
\centerline{
\psfig{file=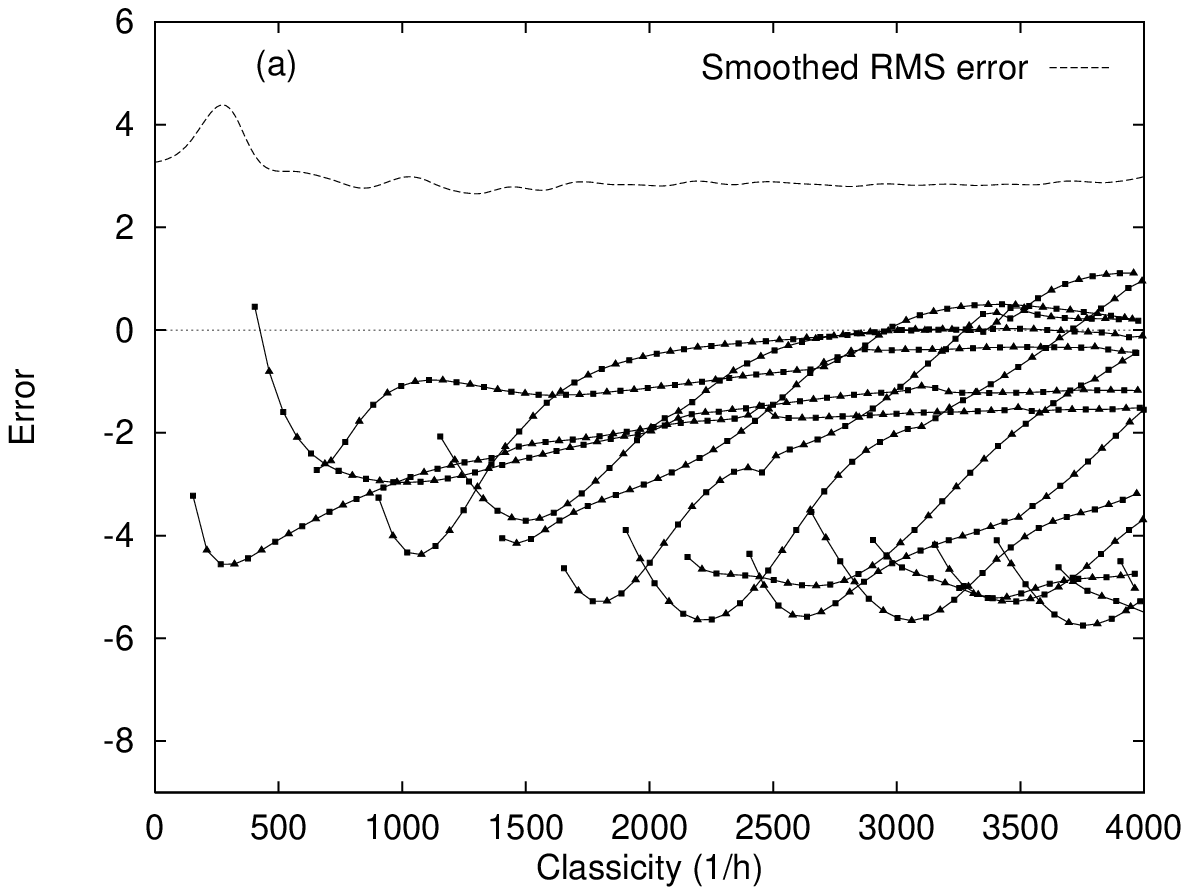,width=12.0cm,height=8.0cm}
}
\centerline{
\psfig{file=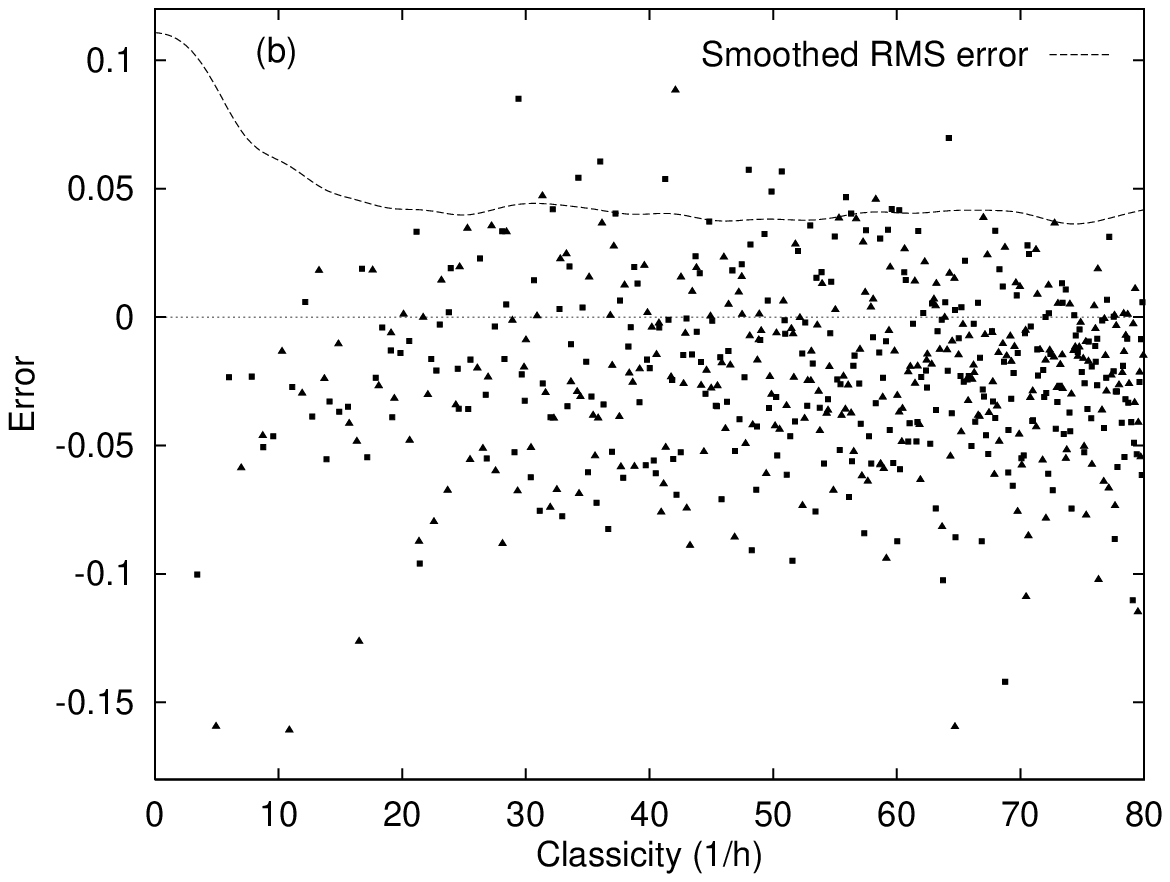,width=12.0cm,height=8.0cm}
}
\caption[Errors in the semiclassical eigenclassicity spectra]{Errors
in the semiclassical eigenclassicity spectra. The discrepancy between
the semiclassically predicted eigenclassicities and the exact values
(obtained by diagonalizing the Hamiltonian), for (a)~$E=0.004$ and
(b)~$E=0.2$. Squares represent even parity states; triangles, odd. In
(a), lines connect eigenstates that are associated with a common
eigenvalue of $T$ and which thus have in common the approximate
quantum number $n'_y$. Corresponding connections are not made in (b)
because in this chaotic regime, associations of eigenstates do not
form anything resembling continuous curves. The dashed curve shows
the RMS error smoothed over small neighborhoods in classicity; note
that it seems to remain roughly constant as the system becomes more
classical.}
\label{fig:errors_abs}
\end{figure}

Figure~\ref{fig:errors_abs} shows the errors of the semiclassical
approximation as a function of classicity, for the two energy values.
Figs.~\ref{fig:errors_abs}(a) and \ref{fig:errors_abs}(b) are scaled
so as to be directly comparable to one another, in the sense that the
vertical and horizontal axes are scaled in proportion to the
respective Thomas-Fermi densities of state for the two energies. Each
symbol on these plots represents an eigenclassicity predicted by
Bogomolny's method; its vertical position shows the amount by which
the semiclassical prediction differed from the exact value. In
Fig.~\ref{fig:errors_abs}(a), line segments connect eigenstates that
are associated with the same $T$-matrix eigenvalue; this was not done
in Fig.~\ref{fig:errors_abs}(b) for the reason mentioned at the end
of the previous section (for that same reason, connecting them would
not result in smooth curves, but rather in a tangled jumble).

We have already discussed some differences between the regular and
the chaotic regimes---that when the system is classically chaotic it
is somewhat more effort to calculate $T$, but somewhat less difficult
to distinguish Class~1 from Class~0 eigenvalues---now we ask: how
accurately does Bogomolny's method predict eigenclassicities in the
two regimes? From our numerical experiment it appears that the
semiclassical method does not care about the degree of classical
chaos; at least in this experiment, {\em eigenstate positions are
approximated by Bogomolny's scheme just as well in the classically
chaotic regime as in the classically regular regime.}

Note also that the worst and RMS average errors seem to be roughly
constant at all classicities---{\em high excitation states' positions
are approximated just as well as low excitation states'.} Moreover,
when one follows individual curves in Fig.~\ref{fig:errors_abs}(a)
(the nearly separable regime), one sees that individually the errors
along any one curve seem to be decreasing towards zero. Remembering
from Section~\ref{sec:qnumbers} that the eigenstates along a given
curve share the same $n_y$ quantum number and have increasing $n_x$
quantum numbers, it seems that, at least in the regular regime, {\em
the semiclassical predictions are better for states which have more
excitations transverse to the surface of section, but roughly
constant regardless of the number of excitations along the SOS.}

Why might this be? We suggest that it is a result of the method's
semi-semiclassical nature: the propogation of a wavefunction
transverse to the surface of section is done semiclassically, and
therefore improves when the number of excitations in that direction
increases. The motion along the SOS, on the other hand, is effected
(conceptually) by matrix multiplications, which are not dependent on
the semiclassical approximation.

One can argue that efforts to find an analogous correlation (between
errors and excitations along or perpendicular to the SOS) in the
classically chaotic regime are doomed to failure because of the lack
of even approximate quantum numbers. Not entirely satisfied by that
argument, we sought such a correlation anyway, but so far without
success.

\begin{figure}
\centerline{
\psfig{file=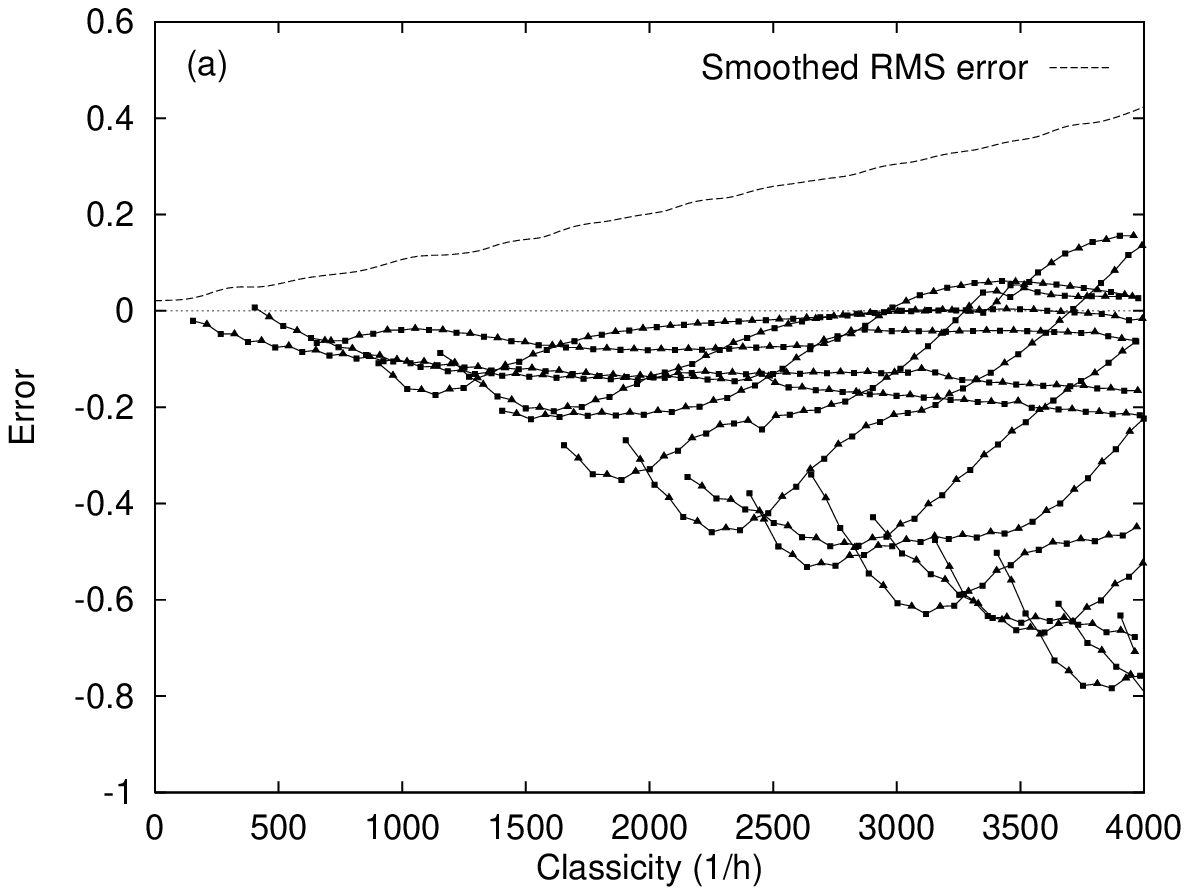,width=12.0cm,height=8.0cm}
}
\centerline{
\psfig{file=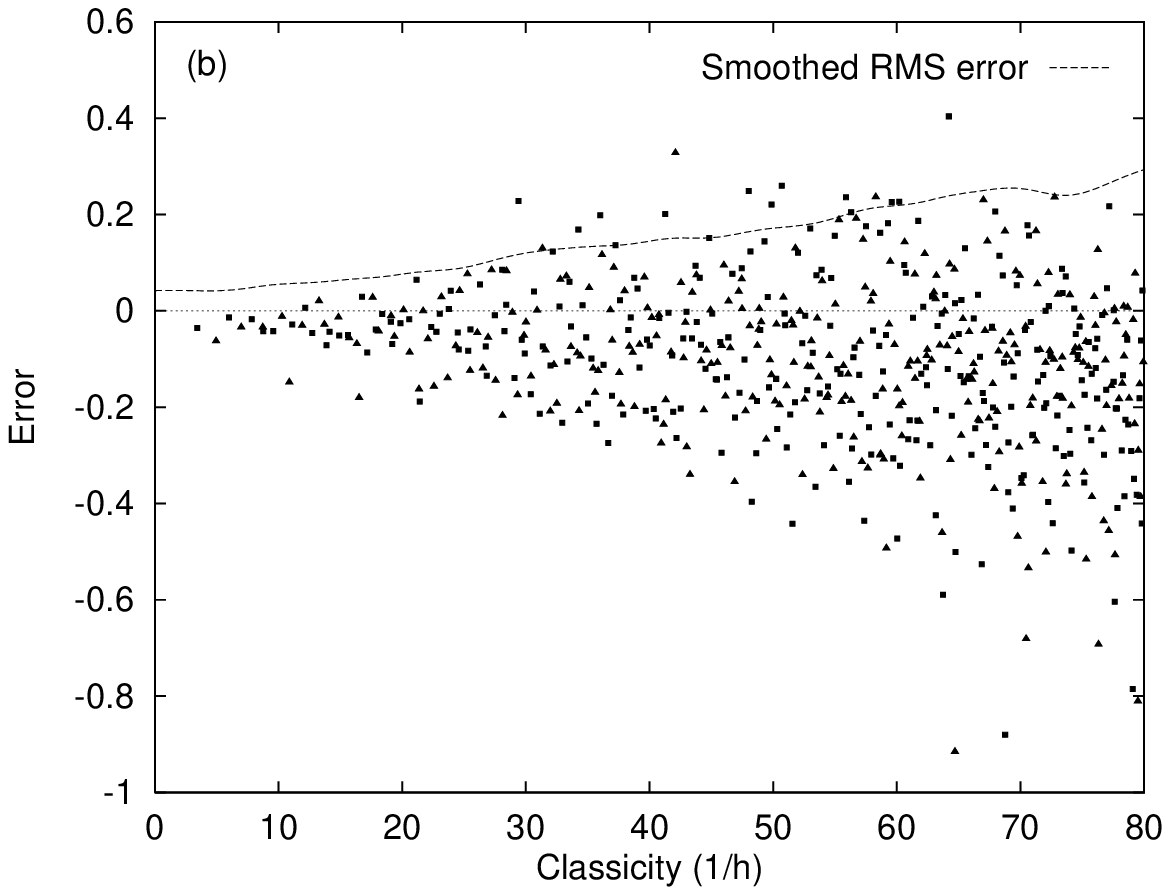,width=12.0cm,height=8.0cm}
}
\caption[Semiclassical errors in units of the mean level
spacing]{Semiclassical errors in units of the mean level spacing.
(a)~$E=0.004$; (b)~$E=0.2$. Alas, the rough constancy in the absolute
errors demonstrated in Fig.~\protect\ref{fig:errors_abs} means that
due to the increasing density of states of the quantum system, the
semiclassical errors {\em as a fraction of the mean level spacing},
shown here, increase with increasing classicity, eventually
preventing the resolution of individual eigenstates semiclassically.
Nevertheless, after predicting nearly 600 eigenstates in each case,
the worst semiclassical errors have yet to quite reach one mean level
spacing, and the RMS averages (the dashed curves) less than half
that.}
\label{fig:errors_frac}
\end{figure}

The above comments refer to the absolute errors (in units of
$1/\hbar$) of Bogomolny's method in approximating the
eigenclassicities of a quantum system. Figure~\ref{fig:errors_frac}
shows to what extent the method is able to meet a more exacting
standard---the ability to resolve individual eigenstates.
There are theoretical reasons to believe that no semiclassical
approximation which is correct only to first order in $\hbar$ will be
able to resolve highly-excited eigenstates of a generic system with
more than one degree of freedom: the density of states increases more
quickly than the semiclassical approximation can hope to converge.
The ability of a method to resolve individual eigenstates is measured
by dividing its errors by the system's {\em mean level spacing}. When
this quantity approaches 1, nearby features of a spectrum can no
longer be separated reliably.

Figure~\ref{fig:errors_frac} shows this ratio for our system. The
mean level spacing decreases like the reciprocal of the classicity,
thereby ``raising the standard'' against which the approximations are
judged. It is seen that Bogomolny's method is a victim of the usual
disease: the ratio of error to desymmetrized level spacing increases
as the classicity is increased, so it will never be able to single
out spectral features at very high excitation number. Still, its
strain of the ailment is relatively nonvirulent---the worst error
ratios are just creeping up towards 1 after hundreds of eigenstates
have been predicted accurately.

\subsection{Calculating surface of section wavefunctions}
\label{sec:SOSwavefunctions}

The eigenstates of the $T$ operator are the values of the quantum
mechanical wavefunction on the surface of section. That is, if
\[ \Psi(x,y) \equiv \langle x,y\vert\Psi\rangle \]
is the 2-D quantum wavefunction, then to within a normalization,
\begin{eqnarray*}
\psi(y) & \equiv & \langle y\vert\psi\rangle \\
& \propto & \Psi(0, y)\;.
\end{eqnarray*}
Odd parity surface of section eigenstates (which are zero on the SOS)
can also be found---intuitively, by moving the surface of section an
infinitesmal distance $\delta x$ from the $y$-axis:
\begin{eqnarray*}
\psi_{\text{odd}}(y) & \propto & \Psi(\delta x, y) \\
& \propto & \left. \frac{\partial \Psi(x, y)}{\partial x} \right|_{x
= 0}\;.
\end{eqnarray*}

As usual, the eigenvalue problem only gives us the semiclassical
wavefunctions to within a complex prefactor. Naturally we choose the
magnitude of this prefactor to normalize the vector to 1, but there
is still a complex phase that needs to be determined.

Since ideally, a phase could be chosen to make the SOS wavefunction
pure real, it is sensible to choose the phase so as to minimize the
imaginary part. Specifically, we try to minimize
\[ I \equiv \int dy \, [{\rm Im}\, \langle y\vert\psi\rangle ]^2. \]
Conveniently, this integral need never be done. Assuming that $\vert
n\rangle$ is an orthogonal complete basis, and that $\langle y\vert
n\rangle$ is always real, the required prefactor phase is simply
\begin{equation}
e^{i\phi}=\pm
\sqrt{
\frac{\sum_n\langle n\vert\psi\rangle^2}
{\vert\sum_n\langle n\vert\psi\rangle^2\vert}}\;.
\label{eq:psi_phase}
\end{equation}
Since we store the SOS eigenfunctions in such a basis, only the sums
appearing in (\ref{eq:psi_phase}) need to be done and no integrals.
It turns out that after this best phase is chosen, the SOS
wavefunctions indeed turn out to have only small imaginary parts.

A sequence of semiclassically predicted surface of section
wavefunctions is plotted in Figs.~\ref{fig:SOSpsi-a} and
\ref{fig:SOSpsi-b}, along with their exact counterparts. In each
energy regime, 6 such plots are presented, at equivalent parts of
their spectra (around the 230th eigenstate, which is about the 115th
eigenstate of a particular parity). The eigenstates were not
specially selected, and are typical of other eigenstates that we
looked at. The three curves plotted in each set are: (1)~the exact
$\psi(y)^2$, which is necessarily real; (2)~the semiclassically
predicted $\vert\psi(y)\vert^2$; and (3)~the square of the residual
imaginary part of the semiclassically predicted wavefunction, $[{\rm
Im}\, \psi(y)]^2$ (which ideally should be zero).

It can be seen from those figures that the semiclassical SOS
wavefunctions capture, in almost all cases, the qualitative features
of their exact counterparts. It is true that the semiclassical
prediction is often rather poor at predicting the relative heights of
peaks in the probability, but its estimate of oscillation length
scales along the SOS are usually quite accurate. It can be seen that
the semiclassical SOS wavefunctions are often too ``sqeezed in'' near
the classical turning points, since they cannot model tunnelling.
However, in many cases the details are predicted with surprising
fidelity.

\newcommand{\gallerypic}[1]{\psfig{file=#1,width=7.5cm}}

\begin{figure}
\centerline{
\hbox{\gallerypic{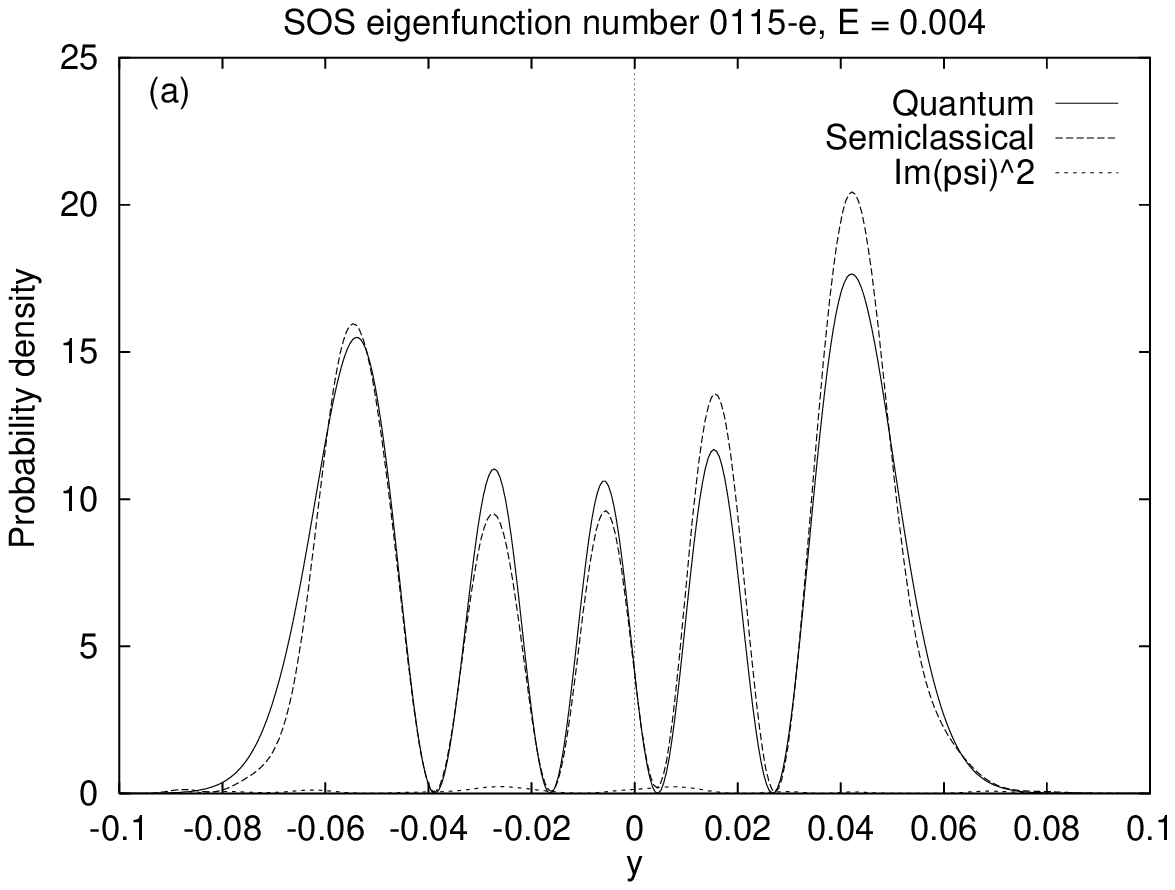}
\gallerypic{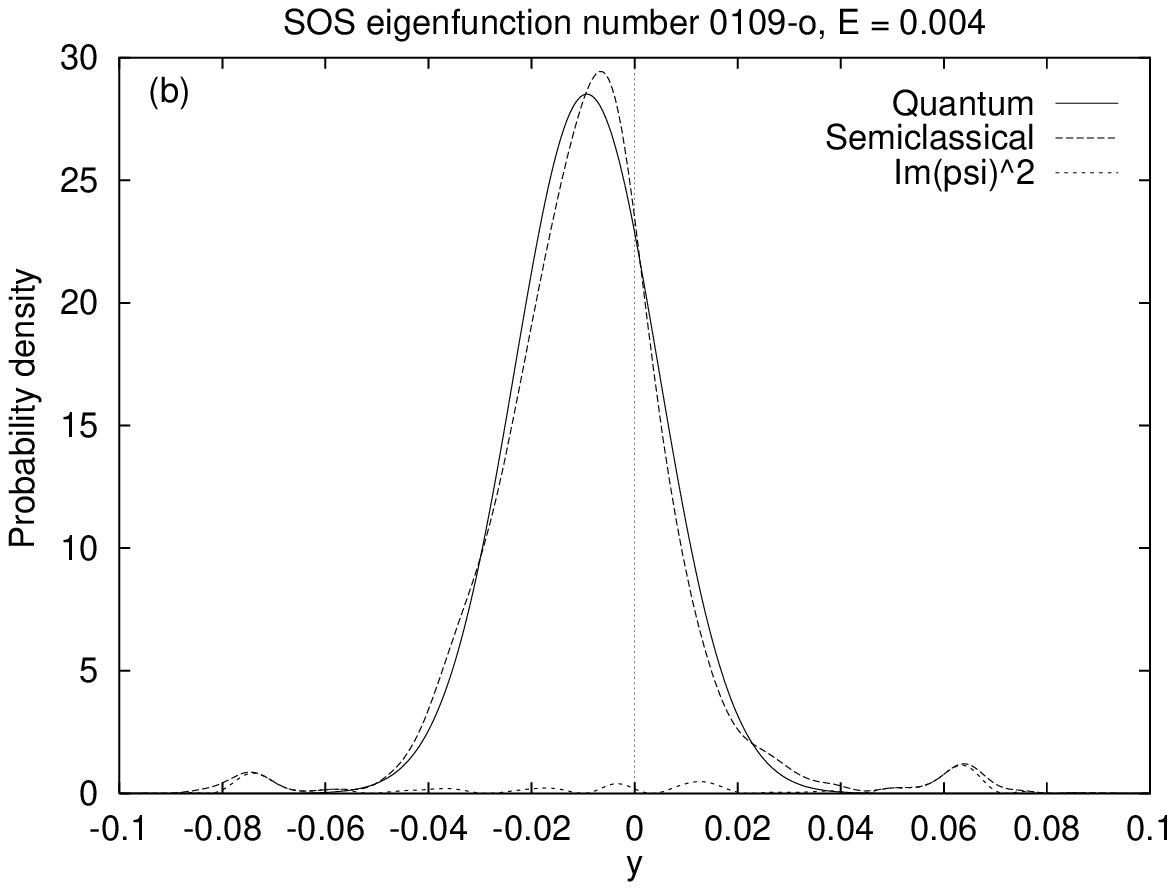}}
}
\centerline{
\hbox{\gallerypic{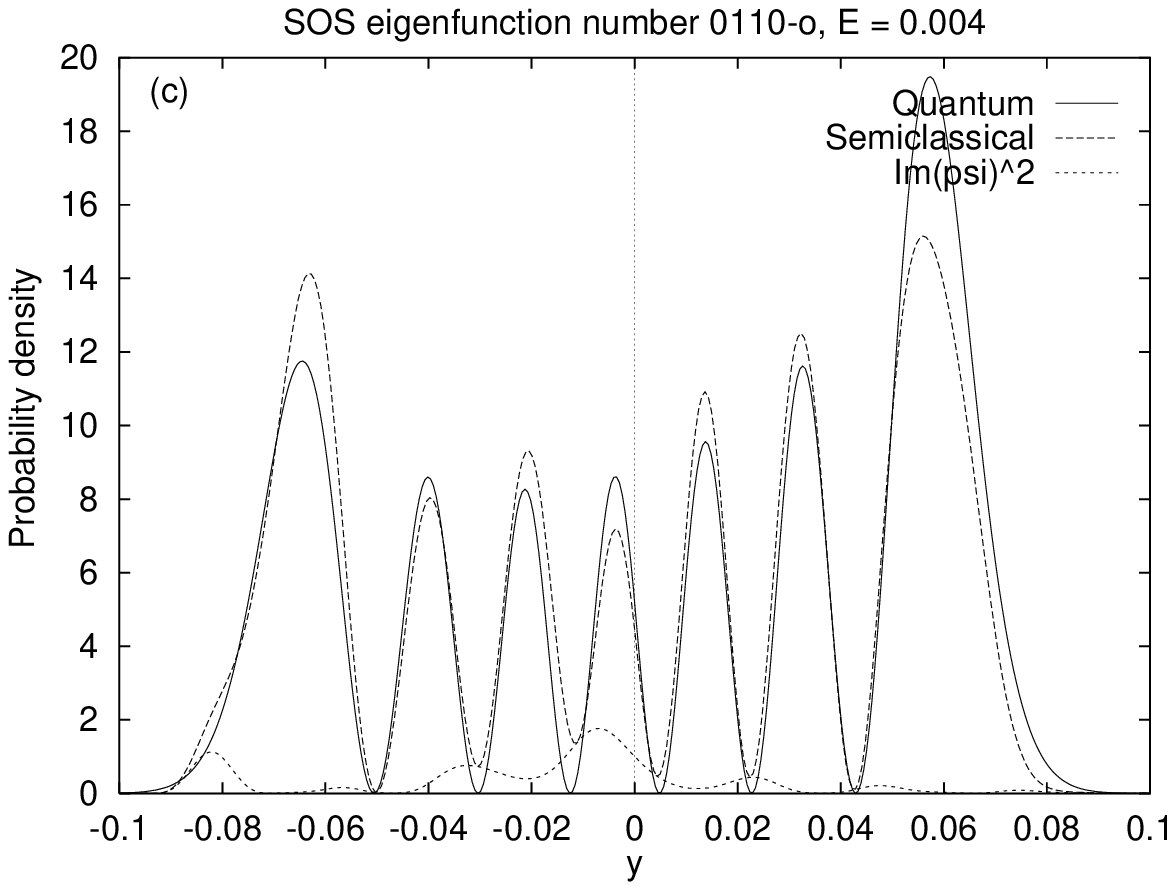}
\gallerypic{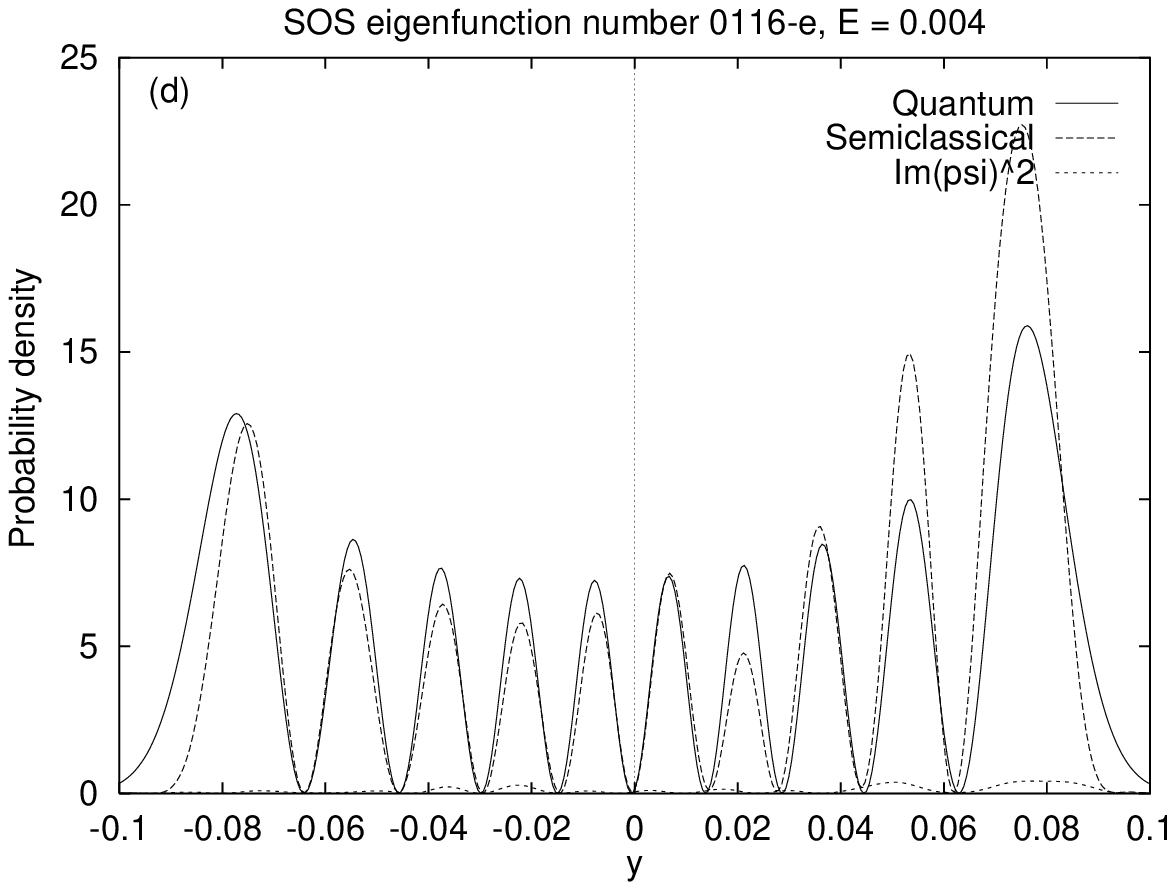}}
}
\centerline{
\hbox{\gallerypic{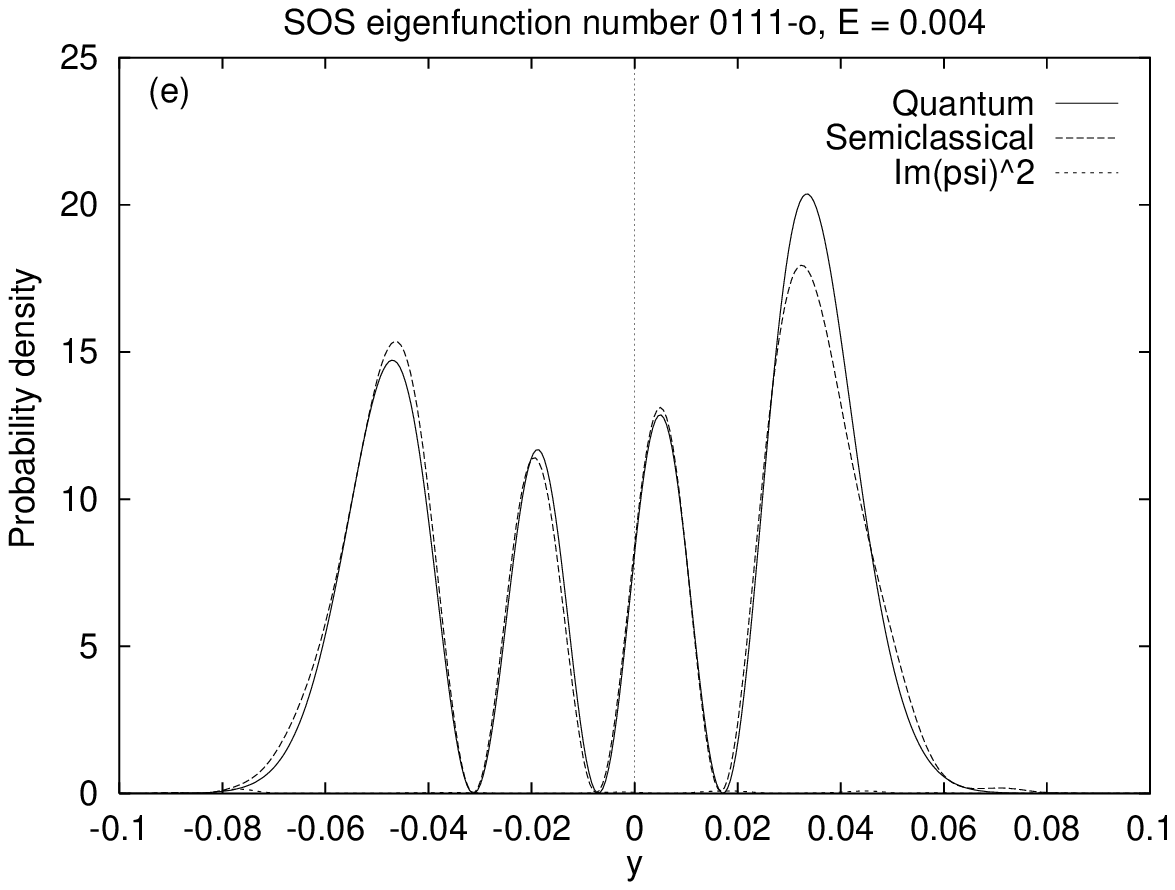}
\gallerypic{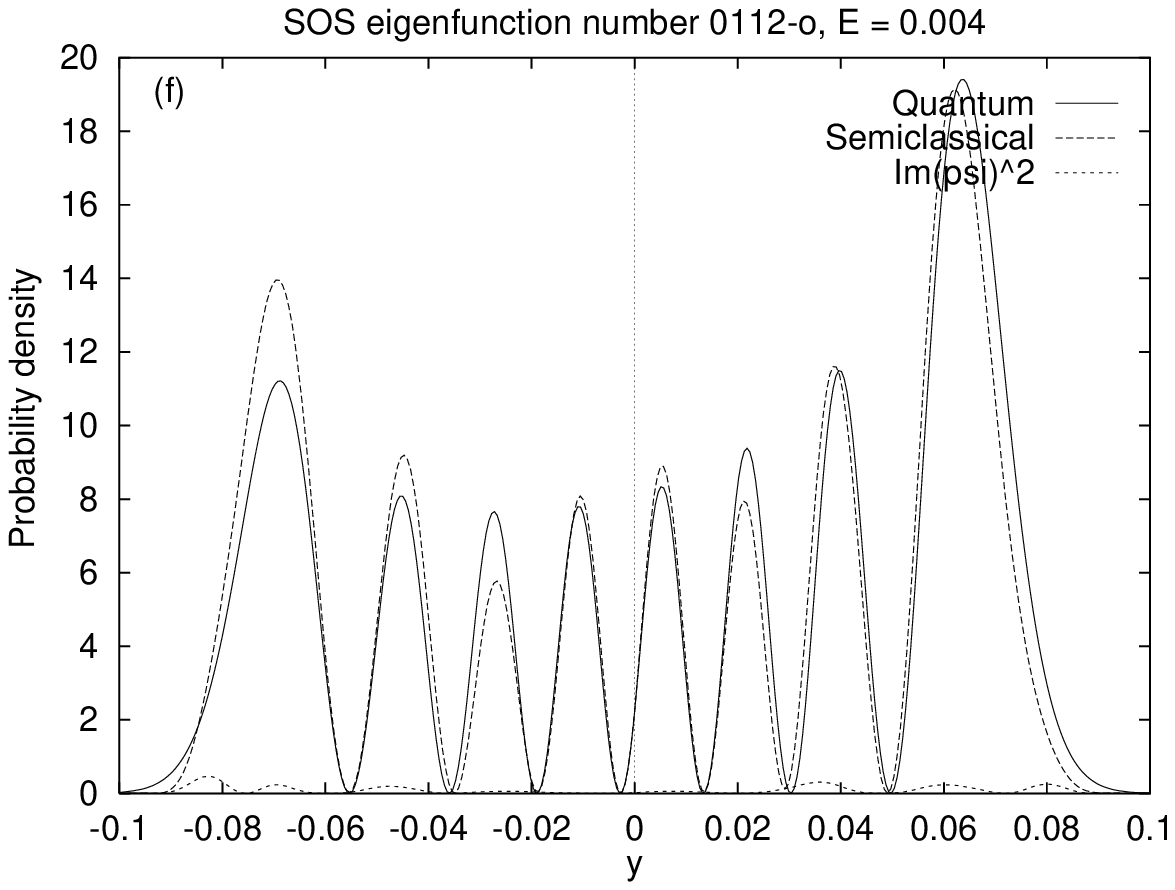}}
}
\caption[SOS eigenfunctions, regular regime]{SOS eigenfunctions,
regular regime. The eigenstates covered are the 6 starting at
classicity $1/\hbar = 2500$.
(a)~115th even eigenstate: exact $1/\hbar=2506.08$;
$\Delta(1/\hbar)=-0.39$.
(b)~109th odd eigenstate: exact $1/\hbar=2508.20$;
$\Delta(1/\hbar)=-1.69$.
(c)~110th odd eigenstate: exact $1/\hbar=2511.33$;
$\Delta(1/\hbar)=-2.38$.
(d)~116th even eigenstate: exact $1/\hbar=2521.05$;
$\Delta(1/\hbar)=-5.26$.
(e)~111th odd eigenstate: exact $1/\hbar=2525.75$;
$\Delta(1/\hbar)=-0.18$.
(f)~112th odd eigenstate: exact $1/\hbar=2537.00$;
$\Delta(1/\hbar)=-4.25$.}
\label{fig:SOSpsi-a}
\end{figure}
\clearpage

\begin{figure}
\centerline{
\hbox{\gallerypic{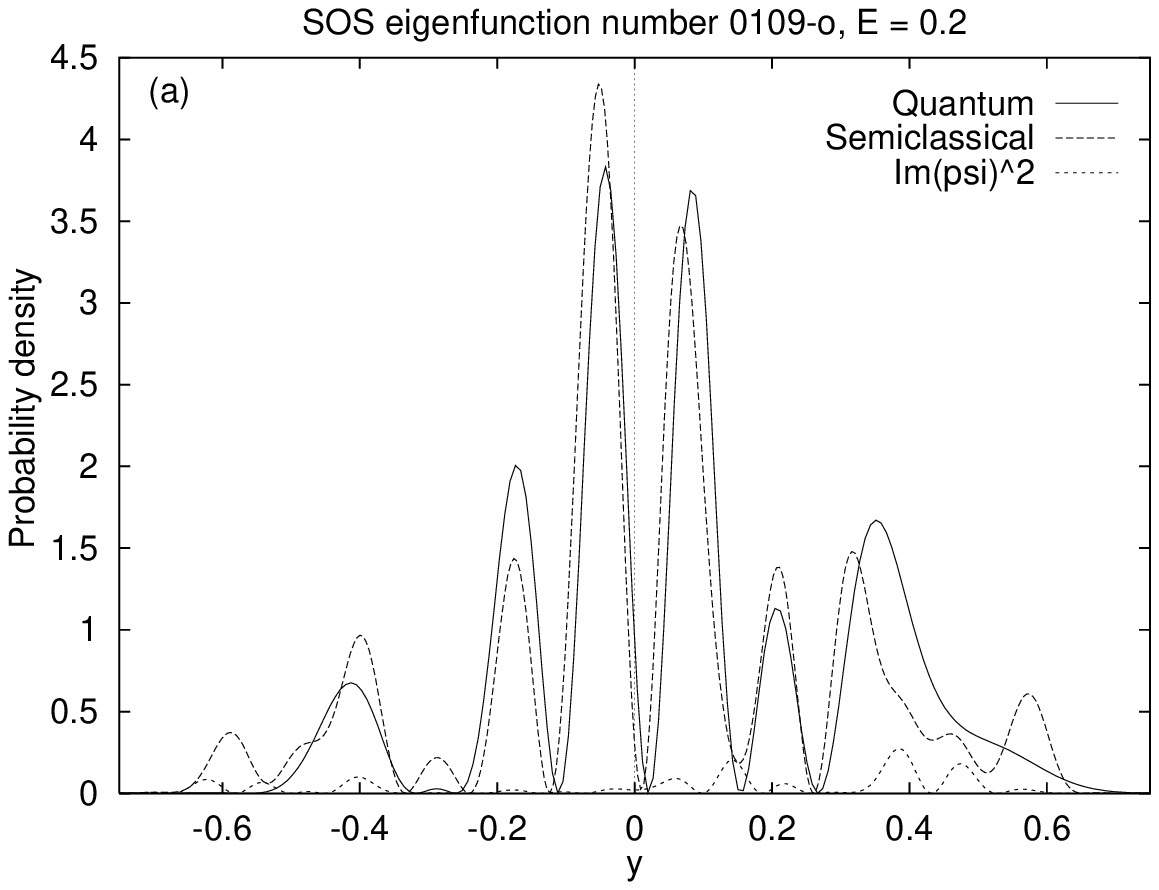}
\gallerypic{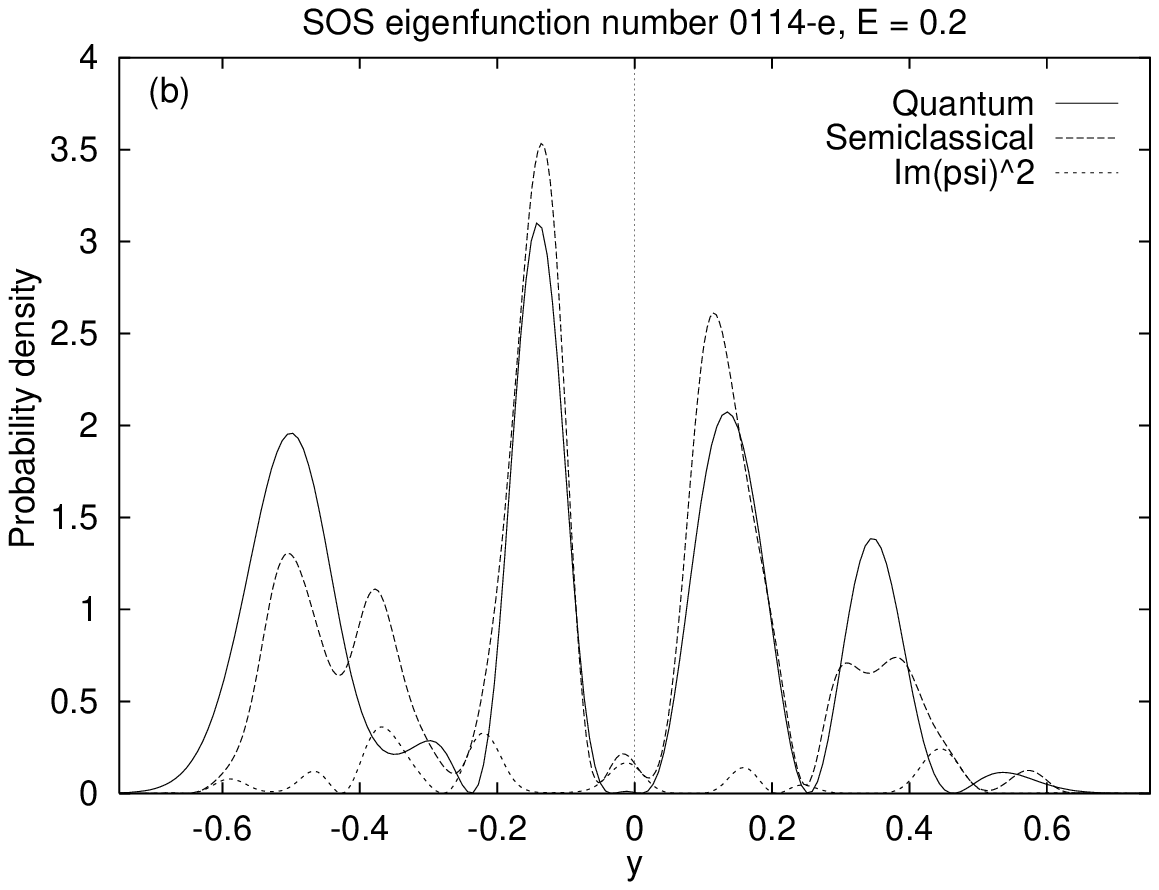}}
}
\centerline{
\hbox{\gallerypic{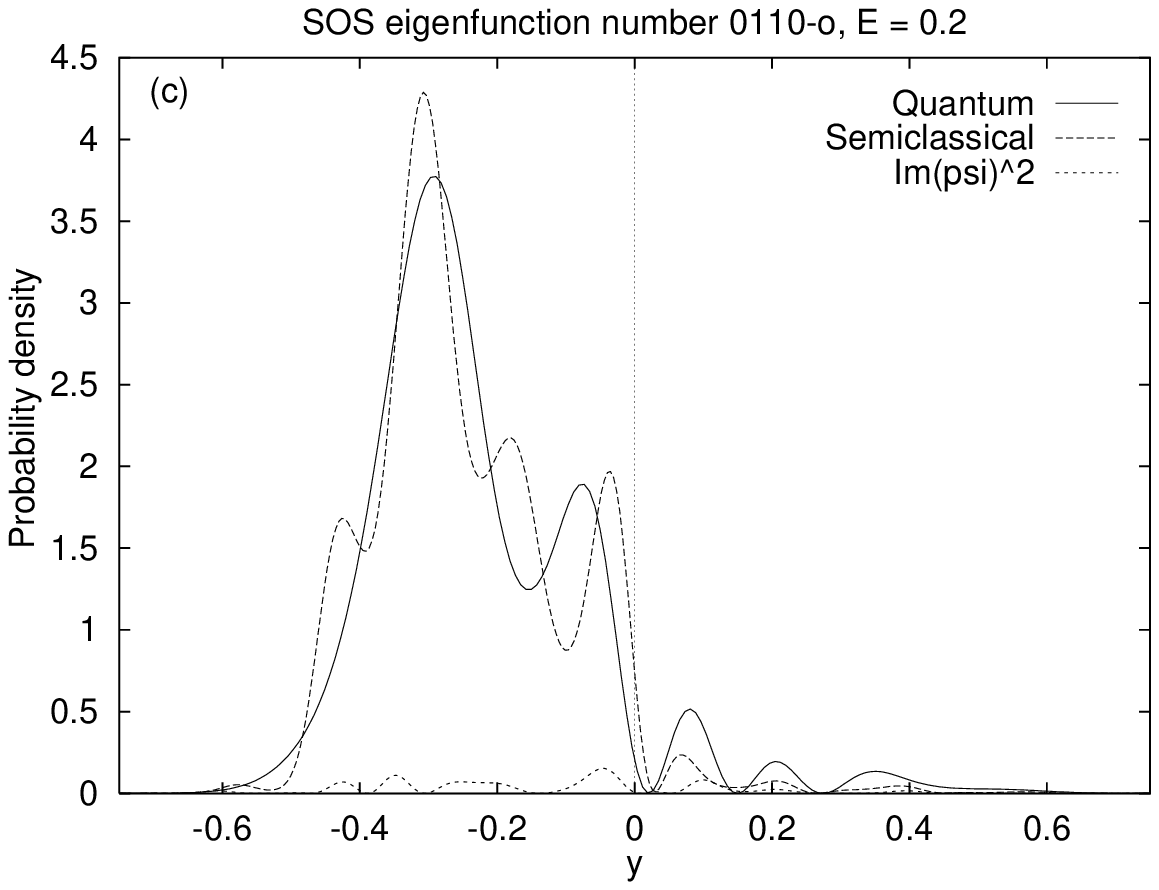}
\gallerypic{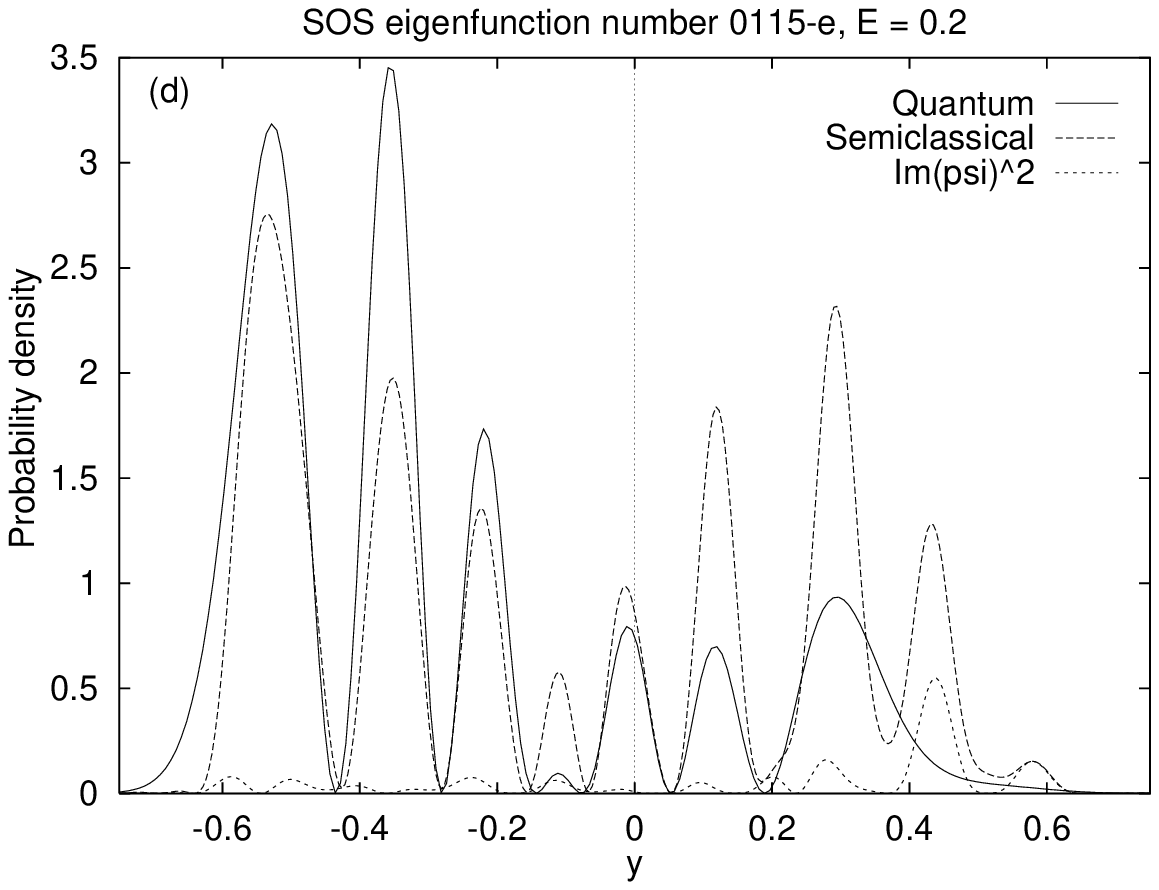}}
}
\centerline{
\hbox{\gallerypic{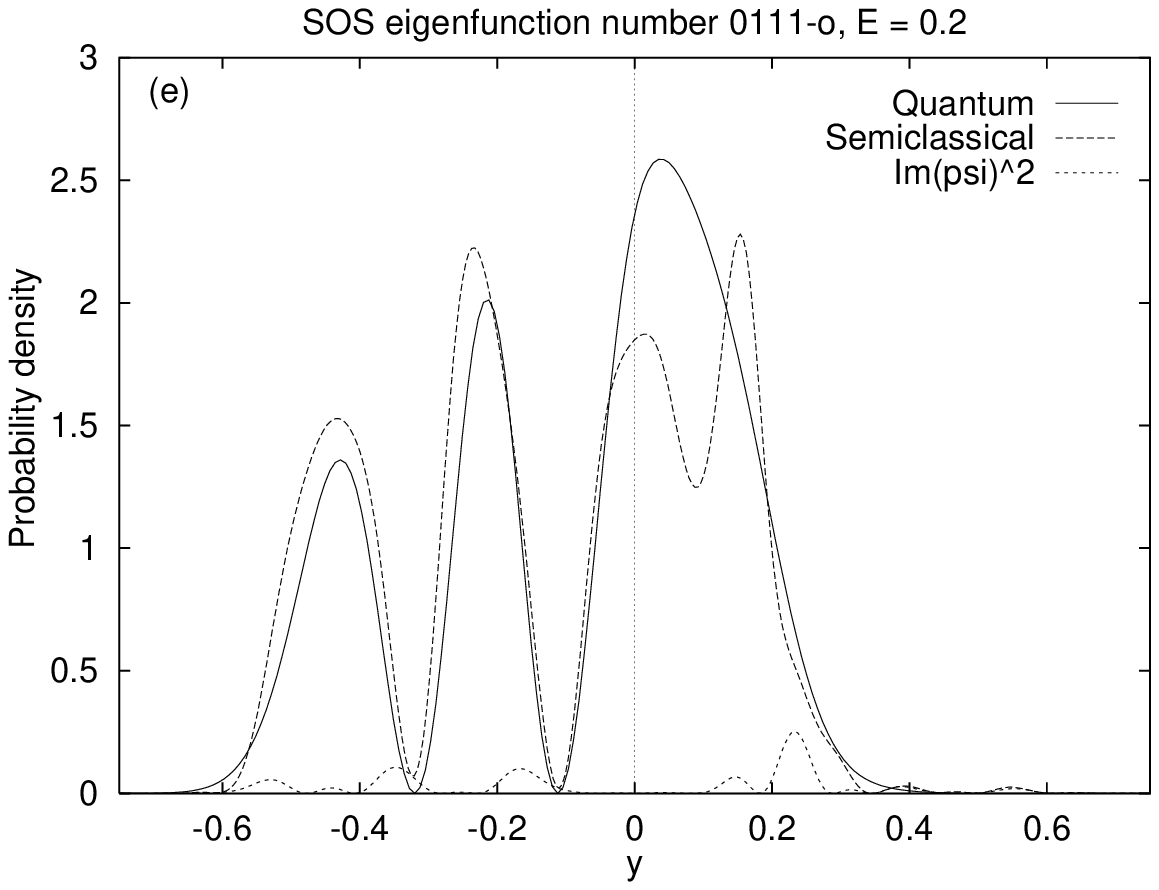}
\gallerypic{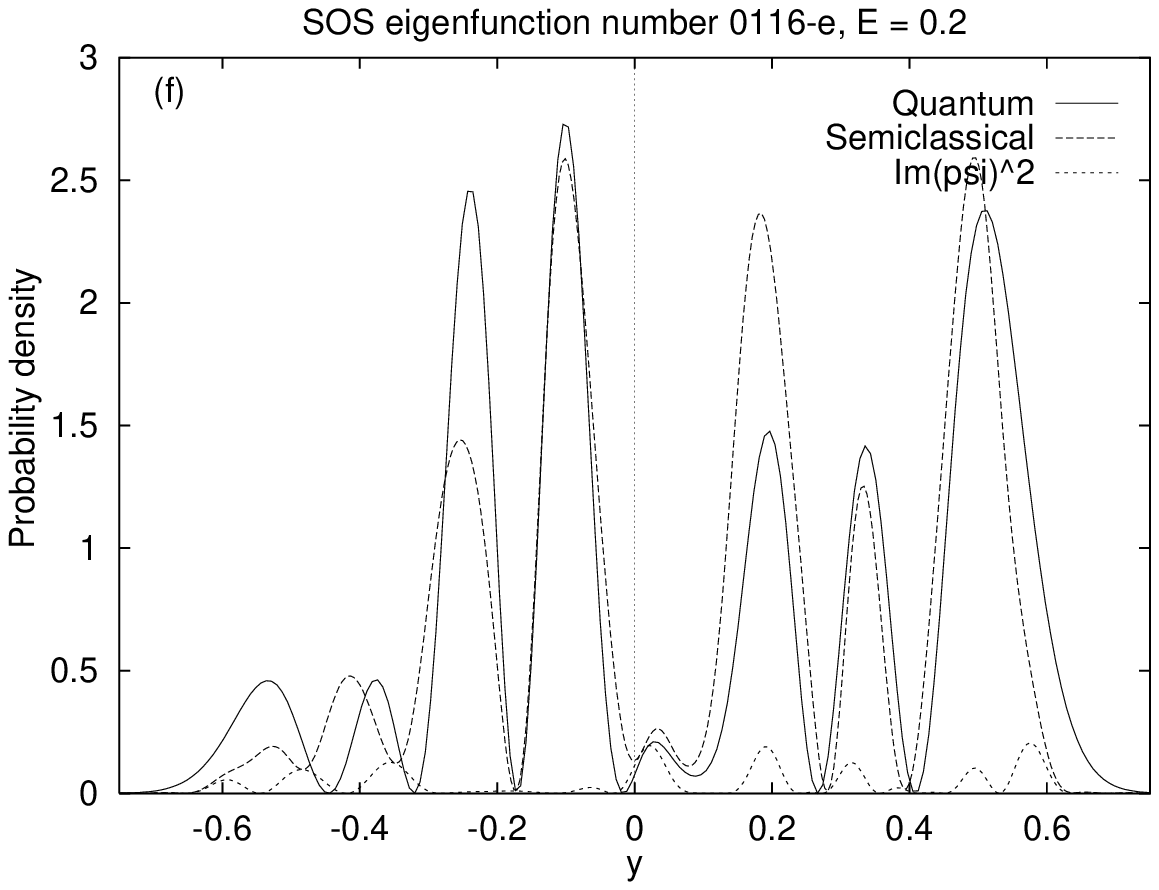}}
}
\caption[SOS eigenfunctions, chaotic regime]{SOS eigenfunctions,
chaotic regime. The eigenstates covered are the 6 starting at
classicity $1/\hbar = 50$.
(a)~109th odd eigenstate: exact $1/\hbar=50.1654$;
$\Delta(1/\hbar)=-0.0421$.
(b)~114th even eigenstate: exact $1/\hbar=50.2430$;
$\Delta(1/\hbar)=-0.0325$.
(c)~110th odd eigenstate: exact $1/\hbar=50.2787$;
$\Delta(1/\hbar)=-0.0060$.
(d)~115th even eigenstate: exact $1/\hbar=50.3252$;
$\Delta(1/\hbar)=-0.0526$.
(e)~111th odd eigenstate: exact $1/\hbar=50.5255$;
$\Delta(1/\hbar)=-0.0418$.
(f)~116th even eigenstate: exact $1/\hbar=50.7136$;
$\Delta(1/\hbar)=0.0534$.}
\label{fig:SOSpsi-b}
\end{figure}
\clearpage

\section{Conclusions}

What is the point of a semiclassical theory? Historically,
semiclassical theory came before, and inspired, matrix mechanics.
However, in the intervening years, as the quantum method revealed its
power and wide applicability, semiclassical methods barely inched
forward. But finally in the late 1960's and early 1970's, two things
happened. One was the (re-)discovery of chaos, and the realization
that a big fraction of classical systems had been left unexplored and
misunderstood, unknowingly assumed nonexistent by scientists whose
training was virtually limited to ballistic trajectories, harmonic
oscillators, and two-body Kepler problems. The second thing that
happened was Gutzwiller's discovery of his periodic orbit theory for
semiclassical quantization. For the first time, semiclassical
mechanics was liberated from the torus. The fashionable blending of
these two developments, dubbed {\em quantum chaos}, is at its core
nothing more than an attempt to understand semiclassical mechanics
off the torus.

Gutzwiller's trace formula is a beautiful edifice, so elegant that
physicists have the gut feeling that it {\em must} be right. This
makes it all the more frustrating that it is so hard to use. Periodic
orbits are wonderful, canonically invariant objects that are easy to
picture and describe. Unfortunately, very long period orbits in a
typical chaotic potential are also furiously difficult to calculate.
A speck of initial conditions, in a moderately to highly chaotic
system, stretches into a gossamer hairball after only a few
oscillations, and finding periodic orbits means finding places where
the hairball and speck coincide---for all possible specks of initial
conditions. A few such heroic computations have been done, and they
are indeed able to reproduce the gross features of the quantum
spectrum, and even individual low-lying states. However, as a
practical method the trace formula has a long way to go.

It is not necessary to discard periodic orbit theory, but maybe it
{\em is} time to expand our toolbox. It has been noted with
admiration that periodic orbits, of longer and longer period,
eventually densely explore every part of the phase space. Thus, it is
argued, when we go to long enough orbits, the periodic orbits will
``know'' all there is to be known about the system's classical
mechanics. But this is vast overkill. We do not need to limit
ourselves to periodic orbits if we want to explore all of phase
space. Any set of trajectories---if sprinkled finely enough---does
the job very nicely.

Bogomolny's quantum surface of section method does just this. It
democratically solicits the contributions of any and every
trajectory. When the vote is over, periodic orbits still have
disproportionate influence. But their influence comes incidentally,
only because periodic orbits come with an entourage of similar
behavior, non-periodic trajectories.

This paper presented an exploration of Bogomolny's method. We
explained how to apply this technique to an arbitrary potential in a
practical way, and estimated just how efficient the method is when
applied to systems of different dimension and different degrees of
chaos. We suggested a practical and general way, by solving for
eigenclassicities rather than eigenenergies, of testing this and
other semiclassical theories with reduced effort. Then we used
Bogomolny's method to perform a semiclassical analysis of a generic,
non-scalable, nonlinear oscillator, giving practical advice and
techniques that will be useful to future users of the method. Our
computation yielded hundreds of eigenvalue predictions in both the
classically regular and the classically chaotic regimes, all accurate
to less than a mean level spacing. We explored some of the properties
of the $T$ operator, especially its dimension and the nature of its
unitarity. Finally, we also computed the surface of section
wavefunctions predicted by the method and found that they also agree
quite well with their exact counterparts.

The hybrid nature of Bogomolny's transfer operator---produced by
summing classical trajectories, but then diagonalized using matrix
methods---makes it something of a {\em semi-}semiclassical method. As
such, it does not represent the yet-unattained ``Holy Grail'' of a
{\em purely semiclassical method} which is able to {\em resolve
arbitrarily highly excited eigenenergies} (indeed, it falls short on
both counts). What this method {\em is}, however, is a practical
method of semiclassically approximating information about quantum
systems; a method which, though somewhat intricate to implement the
first time, can function as a self-contained ``black-box'' that
inputs Hamiltonians and outputs approximate quantum-mechanical
spectra.

\acknowledgements
This work was supported in part by the National Science Foundation,
Grant No. PHY91-15574.

\appendix
\section{Rescaling the Nelson$_2$ Potential}
\label{sec:rescaling}

In this appendix, we discuss the rescaling of the ``Nelson$_2$''
potential that we use in the text, and its connection to the scaling
for the true Nelson potential used by other authors (for example,
\cite{BarDav87}). We also discuss a different way of viewing the act
of varying Planck's constant (as was done in the main text in the
form of the classicity): by adding an additional parameter and
rescaling the dynamical variables, the same effect can be obtained
while using a constant value for $\hbar$.

\subsection{Connection to the ``Nelson'' potential}

The system which has been given the name ``Nelson'' is defined by
\begin{equation}
\bar{H} = {1\over2} \bar{p}_x^2 + {1\over2} \bar{p}_y^2
+ {1\over2} \bar{\mu} \bar{x}^2
+ \left( \bar{y} - {1\over2} \bar{x}^2 \right)^2
\label{eq:Nelson}
\end{equation}
Here, overbars are used to distinguish the variables in this scheme
from our variables. For Nelson$_2$, the analogous equation is:
\begin{equation}
H = {1\over2} p_x^2 + {1\over2} p_y^2
+ {1\over2} \omega^2 x^2
+ {1\over2} \left( y - {1\over2} x^2 \right)^2
\label{eq:Michael}
\end{equation}
The only difference is the factor of ${1\over2}$ preceding the
nonlinear term in the new scaling, added as a minor convenience. The
two sets of dynamical variables are related to one another by a
simple scaling:
\begin{eqnarray*}
q_{x,y} & = & \bar{q}_{x,y} \\
p_{x,y} & = & {1\over\sqrt{2}} \bar{p}_{x,y} \\
H & = & {1\over2} \bar{H} \\
\omega^2 & = & {1\over2} \bar{\mu} \\
t & = & \sqrt{2} \, \bar{t} \\
\hbar & = & {1\over\sqrt{2}} \bar{\hbar} \\
S & = & {1\over\sqrt{2}} \bar{S} \\
\end{eqnarray*}
We should point out that our numerical experiments were done at a
value of $\omega = \sqrt{0.05}$, which is equivalent to the choice
$\bar{\mu} = 0.1$ often used in Nelson potential analyses. Our
selected energies $E = 0.004$ and $E = 0.2$ are equivalent to Nelson
energies $\bar{E} = 0.008$ and $\bar{E} = 0.4$.

\subsection{Making $\hbar$ constant again}

As mentioned in the text, changing $\hbar$ (in the form of the
classicity) is equivalent to a rescaling of the other dynamical
variables. In the following we present the transformation which gives
$\hbar$ back the constant value that Mother Nature intended (namely
1) by inserting a different parameter, $\tilde{\alpha}$, in the
system. The equation which we now wish to compare to
equation~(\ref{eq:Michael}) is as follows:
\begin{equation}
\tilde{H} = {1\over2} \tilde{p_x}^2 + {1\over2} \tilde{p_y}^2
+ {1\over2} \tilde{\omega}^2 \tilde{x}^2
+ {1\over2} \left( \tilde{y} -
{1\over2} \tilde{\alpha}^2 \tilde{x}^2 \right)^2
\label{eq:no_hbar}
\end{equation}
The other difference is that before, $\left[ q_{x,y}, p_{x,y} \right]
= i \hbar$, whereas now, $\left[ \tilde{q}_{x,y}, \tilde{p}_{x,y}
\right] = i$.

Again, a trivial scaling distinguishes the two schemes, so
``changing'' $\hbar$ is equivalent to scaling the other dynamical
variables as follows:
\begin{eqnarray*}
\tilde{q}_{x,y} & = & q_{x,y} / \sqrt{\hbar} \\
\tilde{p}_{x,y} & = & p_{x,y} / \sqrt{\hbar} \\
\tilde{H} & = & H / \hbar \\
\tilde{\omega} & = & \omega \\
\tilde{t} & = & t \\
\tilde{\alpha} & = & \hbar \\
\tilde{\hbar} & = & 1 \\
\tilde{S} & = & S / \hbar \\
\end{eqnarray*}


\begin{thebibliography}{99}
\bibitem{footnote}
Unfortunately, although we used the first optimization, we did not
notice the second one until our computations (in an oscillator basis)
were finished, so updating the $T$ matrix was the most expensive part
of the computational cost---though not prohibitively so.
\end{thebibliography}
\end{document}